\documentclass[twocolumn]{aastex63}
\usepackage{lineno}
\usepackage{tabularx}
\usepackage{xspace}
\usepackage{csvsimple}
\usepackage{rotating} 
\usepackage{comment} 
\usepackage{tabularx}

\newcommand{\ith}{\ensuremath{^{\rm th}}}
\newcommand{\teff}{\ensuremath{T_{\mbox{\scriptsize eff}}}}
\newcommand{\logg}{\ensuremath{\log g}}
\newcommand{\sigmavz}{$\sigma_{v_z}$}
\newcommand{\kms}{\ensuremath{\mbox{km s}^{-1}}}
\newcommand{\prot}{\ensuremath{P_{\mbox{\scriptsize rot}}}}

\submitjournal{\textit{The Astronomical Journal}}

\graphicspath{{./}{figures/}}

\newcommand{\rvar}{$R_{\rm var}$}
\newcommand{\kepler}{\textrm{Kepler} }
\newcommand{\gaia}{\textrm{Gaia} }
\newcommand{\tess}{\textrm{TESS}}
\newcommand{\ro}{$R_o$\xspace}

\shorttitle{Gyro-Kinematic Ages for \kepler Stars}
\shortauthors{Lu et al.}

\begin{document}

\title{Gyro-Kinematic Ages for around 30,000 \kepler Stars}

\correspondingauthor{Yuxi (Lucy) Lu}
\email{lucylulu12311@gmail.com}

\newcommand{\amnh}{American Museum of Natural History, Central Park West, Manhattan, NY, USA}
\newcommand{\cca}{Center for Computational Astrophysics, Flatiron Institute, 162 5\ith\ Avenue, Manhattan, NY, USA}
\newcommand{\columbia}{Department of Astronomy, Columbia University, 550 West 120\ith\ Street, New York, NY, USA}

\author[0000-0003-4769-3273]{Yuxi(Lucy) Lu}
\affiliation{\columbia}
\affiliation{\amnh}

\author[0000-0003-4540-5661]{Ruth Angus}
\affiliation{\amnh}
\affiliation{\cca}
\affiliation{\columbia}

\author[0000-0002-2792-134X]{Jason L.\ Curtis}
\affiliation{\amnh}

\author[0000-0001-6534-6246]{Trevor J.\ David}
\affiliation{\amnh}
\affiliation{\cca}

\author[0000-0003-2102-3159]{Rocio Kiman}
\affiliation{\amnh}
\affiliation{Department of Physics, Graduate Center, City University of New York, 365 5th Ave, New York, NY 10016, USA}
\affiliation{Department of Physics and Astronomy, Hunter College, City University of New York, 695 Park Avenue, New York, NY 10065, USA}

\begin{abstract}
Estimating stellar ages is important for advancing our understanding of stellar and exoplanet evolution and investigating the history of the Milky Way.
However, ages for low-mass stars are hard to infer as they evolve slowly on the main sequence.
In addition, empirical dating methods are difficult to calibrate for low-mass stars as they are faint.
In this work, we calculate ages for Kepler F, G, and crucially K and M dwarfs, using their rotation and kinematic properties.
We apply the simple assumption that the velocity dispersion of stars increases over time and adopt an age--velocity--dispersion relation (AVR) to estimate average stellar ages for groupings of coeval stars.
We calculate the vertical velocity dispersion of stars in bins of absolute magnitude, temperature, rotation period, and Rossby number and then convert velocity dispersion to kinematic age via an AVR.
Using this method, we estimate \textit{gyro-kinematic} ages for 29,949 Kepler stars with measured rotation periods.
We are able to estimate ages for clusters and asteroseismic stars with an RMS of 1.22 Gyr and 0.26 Gyr respectively.
With our \texttt{Astraea} machine learning algorithm, which predicts rotation periods, 
we suggest a new selection criterion (a weight of 0.15) to increase the size of the \citet{McQuillan2014} catalog of Kepler rotation periods by up to 25\%.
Using predicted rotation periods, we estimated gyro-kinematic ages for stars without measured rotation periods and found promising results by comparing 12 detailed age--element abundance trends with literature values.

\end{abstract}


\section{Introduction} \label{sec:intro}
The age of a star is one of its most important, yet difficult to determine, quantities.
Stellar ages are useful in many fields and at many geometric scales.
On a small scale, the age of exoplanets can be inferred from their host stars, and can be used to study how the properties of exoplanets change over time \citep[e.g.,][]{David2020}.
On the galactic scale, the ages of stars can be used to understand how the Milky Way formed and is evolving \citep{Spina2018,Bedell2018,Ness2019}.
However, age is not a directly measurable physical quantity, but rather an estimation of a star's evolutionary state.
The main source of uncertainty comes from the fact that the observable features of stars---their luminosities and temperatures---change very slowly while on the main sequence.
This is particularly true of low-mass K and M dwarfs.
As a result, even when measured precisely, these observables cannot tightly constrain the ages of most main-sequence stars.
In general, most available dating methods cannot estimate ages for M dwarfs \citep[for a detailed review of these methods and their application areas, see][]{Soderblom2010}. 
These low-mass stars are faint, which makes observations hard, and they evolve extremely slowly on the main sequence.
In addition, stellar evolution models are often poorly calibrated for these stars.
The M dwarfs with the most accurate and precise age measurements are those in open clusters, or with a binary companion that can be independently dated.
However, this only applies to a small number of stars, and getting precise and accurate ages for old field M dwarfs remains challenging.
\footnote{Some progress for a small amount of M dwarf with high resolution spectroscopic data has been made using chemo-kinematic ages \citep{Veyette2018}, where they estimated M dwarf ages between 4-9 Gyr with uncertainties between 2-3.5 Gyr.}. 

Isochrone fitting is currently the most productive method to infer ages for individual field stars \citep[e.g.,][]{Nordstrom2004,Buder2019,Berger2020}. 
Although it works well on open clusters that contain a population of stars of the same age, this method has trouble estimating precise ages for individual stars.
The density of the isochrones around the zero-age main sequence is extremely high as stars evolve slowly on the main sequence, so this degeneracy of the model prevents isochrone fitting from estimating accurate and precise ages for stars that have not yet gone through $\sim$1/3 of their main-sequence lifetimes \cite[e.g.,][]{Soderblom2010}. 
In other words, the temperatures and luminosities of low-mass stars change extremely slowly on the main sequence and, for this reason, isochrone fitting rarely provides precise ages for K and M dwarfs.
Isochrone ages for K and M dwarf stars can often have uncertainties that are more than 50\%.
Thus, isochrone fitting is not a precise method for getting ages for individual field stars, especially low-mass stars, due to their long main-sequence lifetimes.

One other technique that can potentially be used to age-date low mass, main-sequence field stars is \textit{gyrochronology} \citep[][]{Barnes2003, Barnes2007}. 
Stars spin down over time due to the loss of angular momentum from magnetized winds \citep[e.g.,][]{Kawaler1988,Weber1967,Schatzman1962}.
As a result, the rotation period of a star can be used to probe its age \citep{Skumanich1972}.
Empirical gyrochronology makes the simple assumption that the rotation period and age of a star are correlated and it requires nearly no assumptions about the physics involved. 
One can then estimate the age of a star as long as it has a measured rotation period, and temperature, mass, or color.
Alternatively, there are physics-based semi-empirical methods \citep[e.g.,][]{matt2015, vansanders2013}.
Such an approach has been used to age-date cool dwarf stars with median statistical uncertainties on the order of 10\% \citep{Claytor2020}, although large systematic uncertainties persist in all gyrochronology models  \citep[e.g.,][]{Curtis2019, Curtis2020}.
All empirical and semi-empirical methods need to be calibrated with data for benchmark stars.
Such benchmarks include members of open clusters, individual stars characterized with asteroseismology or interferometry, and binary systems.
Unfortunately, the gyrochronology relations remain uncalibrated for M dwarfs due to the difficulties in getting ages for suitable benchmarks.

Huge efforts and progress have been made to calibrate an empirical gyrochronology relation for FGK stars \citep[e.g.,][]{Barnes2007,Mamajek2008,Meibom2009,Barnes2010,Garcia2014,matt2015,Angus2015,vansaders2016,Curtis2020}.
Large surveys such as \kepler \citep{Borucki2010}, \gaia \citep{Prusti2016,Brown2018} and \tess\ \citep{TESS} are expanding the calibration sample to an extraordinary level;
however, calibrating gyrochronology for M dwarfs with these samples will remain challenging for the following reasons:
\begin{itemize}
    \item {\it Open clusters} 
    are generally young because they dissolve in the Milky Way on a timescale of $\sim$200 Myr \citep{Soderblom2010,Janes1994,Janes1988}. 
    For this reason, older open clusters are rare and few are close and 
    bright enough to provide rotation period measurements for their M dwarf members.
    As a result, the oldest M dwarfs in clusters with measured rotation periods are 
    around 700 Myr old \citep[in Praesepe and the Hyades; e.g.,][]{Rebull2017,Douglas2017, Douglas2019}.\footnote{Although \citet{Agueros2018} did contribute a few early M dwarfs in the 1.4~Gyr NGC 752 cluster, the mid-to-late M dwarfs remained out of reach.}
    Therefore, open clusters cannot yet provide the precise ages and rotation periods for old M dwarfs needed to fully calibrate gyrochronology (at least not with the current generation of telescopes).
    
    \item {\it Asteroseismic stars} with measurable asteroseismic signals are generally massive and/or evolved. 
    It is extremely hard to detect asteroseismic signals in low-mass stars that are on the main sequence mostly due to the low amplitudes \citep[e.g.,][]{Rodriguez2016}.
    While asteroseismology can be useful for calibrating gyrochronology for F and G dwarfs and subgiants, it cannot yet provide precise ages for M dwarfs.
    
    \item {\it Binary systems} where the age of one of the stars is determined by other methods can be useful benchmarks. For example, binaries where one star can be age-dated with asteroseismology or white dwarf cooling show promise for calibrating gyrochronology. However, such systems are relatively rare. 
\end{itemize}

For these reasons, calibrating the period--age relation for old, very low-mass field stars on a large scale is challenging.
Fortunately, Galactic kinematics provide another useful indicator of stellar age, which is generally applicable to stars with a range of masses and evolutionary stages.

The vertical velocity dispersion of stars increases over time. 
\cite{Spitzer1951} proposed that older stars have a higher velocity dispersion caused by secular gravitational interactions between these stars and gas clouds.
This theory is supported by subsequent observational studies \citep[e.g.,][]{sellwood2014, Lacey1984, Barbanis1967, Sellwood1984}.
An alternative theory is that these older stars were born kinematically \textit{hot} in the first place \citep{Bird2013}.
It is still debatable which theory is correct but it is possible that both of these mechanisms play significant roles in shaping the kinematic of the stars \citep[e.g.,][]{Binks2020}.
There is a long history of studying the kinematic properties of stars in the solar neighborhood;
for example, \citet{Stromberg1946} analyzed the velocity of 444 stars within 20 pc of the Sun and found stars with highest galactic latitude have the highest velocity dispersion.
Further observations \citep[e.g.,][]{Nordstrom2004,Holmberg2007,Holmberg2009,Aumer2009,Yu2018,Ting2019} confirmed older stars exhibit higher velocity dispersions, especially in the vertical direction relative to the galactic plane.
This relation connecting the age and velocity dispersion is called the age--velocity dispersion relation (AVR).

We can group stars with similar stellar properties (e.g., temperature and rotation period, which we assume means they also have similar ages), and calculate the velocity dispersion within each group.
Applying an AVR, we can then estimate the kinematic age of a group of stars based on its vertical velocity dispersion. 
We refer to this approach as \textit{gyro-kinematic} age-dating.

One of the advantages of using kinematics to determine ages for stars is the fact that the underlying physics is simple:
kinematic ages are independent of any stellar physics and only rely on the simple assumption that the vertical velocity dispersion of stars increase over time.
A specific AVR does have to be adopted, but \cite{Martig2014} suggested the disk heating mechanism would result in a simple power-law relation for the AVRs so the impact of this model choice and any bias that it introduces is simple and relatively easy to unpack.
As a result, kinematic ages are useful for calibrating other empirical age-dating methods and providing insights into stellar physics.

However, there are caveats to using this method to calibrate empirical relations like gyrochronology. 
Firstly, kinematic ages are averages for a population of stars. 
Instead of estimating the true individual ages, we are inferring the expected ages based on the kinematic properties of likely-coeval stars.
As a result, kinematic ages cannot be calculated for individual stars this way.
Moreover, other sample selection biases could also affect the ages.
For example, the orientation of the \kepler field would favor stars closer to the galactic plane since \kepler observed nearby stars at a low galactic latitude.
This could potentially introduce bias in our age measurement.
Despite this limitation, kinematic ages can still be used to calibrate gyrochronology \citep{Angus2020_kinage} and provide another perspective to study the physics driving stellar evolution.

Here, we present a catalog of gyro-kinematic ages for 29,949 \kepler stars with rotation periods measured by  \cite{McQuillan2014}, \cite{Garcia2014}, and \cite{Santos2019}.
The column description for the gyro-kinematic ages catalog with 29,949 stars with measured rotation periods is shown in table~\ref{tab:kinagetab}.

\begin{table*}
\centering
\caption{Catalog description of the 29,949 gyro-kinematic ages with measured rotation periods.}
\begin{tabular}{crcrX}
\hline
\hline
Column & Unit & Description\\
\hline
\texttt{kepid} & & \kepler ID\\
\texttt{Prot} & days & rotation period \\
\texttt{Prot\_err}& days & error on rotation period \\
\texttt{source\_id}& & Gaia  DR2 Source ID\\
\texttt{ra}& deg & right ascension from Gaia DR2\\
\texttt{ra\_error}& deg & error on right ascension from Gaia DR2\\
\texttt{dec}& deg & declination from Gaia\\
\texttt{dec\_error}& deg & error on declination from Gaia DR2\\
\texttt{all\_vz}& \kms & vertical velocity from radial velocity or Angus et al. (in prep)\\
\texttt{vz\_err\_all}& \kms & error on vertical velocities\\
\texttt{vel\_dis}& \kms & velocity dispersion\\
\texttt{vel\_dis\_err}& \kms & uncertainty on velocity dispersion due to uncertainties on stellar parameters\\
\texttt{kin\_age}& Gyr & gyro-kinematic ages\\
\texttt{kin\_age\_err}& Gyr & error on gyro-kinematic ages combining error from AVR fits and unvcertainties on the stellar parameters\\
\texttt{Ro}& & Rossby number\\
\texttt{teff}& K & temperature\\
\texttt{abs\_G}& mag & absolute magnitude from Gaia DR2\\
\hline
\end{tabular}
\tablecomments{This table is published in its entirety in a machine-readable format in the online journal.}
\label{tab:kinagetab}
\end{table*}

We also investigate the potential for estimating ages for \kepler stars with predicted rotation periods using \texttt{Astraea} \citep{Lu2020}.\footnote{The code is available at \url{https://github.com/lyx12311/Astraea}}
\texttt{Astraea} is a tool to predict rotation periods from stellar properties such as luminosity, temperature and \rvar.
With these predicted periods, we then estimate gyro-kinematic ages and calculate the individual age--element abundance trends for solar twins.

In Section \ref{sec:datamethod}, we describe the data and method used to estimate gyro-kinematic ages.
In Section \ref{sec:results}, we compare the gyro-kinematic ages with isochrone ages from \cite{Berger2020}, M dwarfs in white dwarf binary systems \citep{Kiman2020}, asteroseismic ages \citep{Silva2017}, and open cluster ages \citep{Curtis2020}.
In Section \ref{subsec:limit}, we discuss the limitations and discrepancies between gyro-kinematic ages and ages estimated by other methods.
We also investigate the age--abundance trends found for solar twins in Section \ref{subsec:abund}, using ages from predicted periods and abundances from APOGEE DR16 \citep{Majewski2017, DR162020}, and we compare our results to those of \cite{Bedell2018}.
We conclude in Section \ref{sec:concl}.

\section{Data \& Methods} \label{sec:datamethod}
\subsection{Data}\label{subsec:data}
To construct the gyro-kinematic age catalog, we started with the rotation period catalog from \cite{McQuillan2014} with 34,030 measured rotation periods and added an extra 4,637 stars from \cite{Garcia2014} and \cite{Santos2019}, which, combined, contain 38,667 stars.
We then accessed their \gaia\ data from the publicly available \textit{Kepler}--\textit{Gaia} DR2 cross-matched catalog produced with a 1$''$ search radius.\footnote{Available at \url{gaia-kepler.fun}}

We calculated effective temperatures (\teff) and absolute Gaia $G$-band magnitudes ($M_G$) from these data. 
First, we accounted for reddening and extinction from interstellar dust for each star using the Bayestar dust map implemented in the \texttt{dustmaps} Python package \citep{Green2018}, and \texttt{astropy} \citep{astropy:2013,astropy:2018}.
We then used Gaia DR2 photometric color, $(G_{\rm BP} - G_{\rm RP})_0$, to estimate effective temperatures for the stars in our sample, using the calibrated relation in \cite{Curtis2020}.
We calculated the Rossby number (\ro) using convective overturn time ($\tau$) defined in equation 11 in \cite{Wright2011} and \ro=\prot/$\tau$.

We cross-matched our rotation sample with the spectroscopic catalog produced by the
Large Sky Area Multi-Object Fibre Spectroscopic Telescope \citep[LAMOST;][]{LAMOST} to obtain radial velocity (RV) measurements.
14,328 stars in our sample had RVs from either LAMOST DR5 \cite{Cui2012,Xiang2019} or Gaia DR2.
For stars with RVs, we calculated the vertical velocity, v$_{z}$, using {\tt astropy} \citep{astropy:2013, astropy:2018}.
However, for stars without RV measurements, we inferred their vertical velocities from their Gaia proper motions and 3D positions.
Because stars in the Kepler field are located at low Galactic latitudes, their vertical velocities can be inferred precisely from their proper motions with a typical precision of around 10\% (Angus et al, in prep).
We excluded 5,379 stars hotter than 8000 K and stars with $|$v$_z| >$ 100~\kms.
This left us with 29,949 stars with measured rotation periods and vertical velocities.

In addition to the primary photometric, spectroscopic, and kinematic data from Gaia, LAMOST, and Kepler, we also used multiple age catalogs in order to test our results.
We used the isochrone age catalog from \cite{Berger2020}, the Kepler LEGACY asteroseismic age catalog from \cite{Silva2017}, the open cluster catalog from \cite{Curtis2020}, and white dwarf cooling ages for two white dwarf--M dwarf binaries from \cite{Kiman2020}, calculated using \texttt{wdwarfdate}.\footnote{Available at \url{https://github.com/rkiman/wdwarfdate}}
Individual element abundances from APOGEE DR16 \citep{Majewski2017, DR162020} were also used to compare the individual element abundance trends with stellar age found for solar twins \citep{Bedell2018}.

\subsection{Methods}\label{subsec:method}
To obtain the gyro-kinematic ages for these 29,949 stars, we calculated the vertical velocity dispersion (\sigmavz) for stars in bins of $M_G$, \teff, rotation period (\prot), and Rossby number (\ro), which we refer to as MTPR from now on.
For each star, we calculated \sigmavz\ by taking the median absolute deviation (MAD) of vertical velocities of a group of stars that are close to that star in MTPR space.
We then multiply the MAD by 1.5 to approximate the standard deviation of vertical velocities, assuming the underlying distribution is Gaussian \cite{Rousseeuw1993}. We opted for MAD as it reduces sensitivity to outliers compared to the standard deviation.

We then converted \sigmavz\ to gyro-kinematic age using the AVR for metal-rich stars ([Fe/H] $> -0.2$ dex) from \cite{Yu2018}.
This AVR was calibrated using 3,564 subgiants and red giants ($5000 < \teff < 5300$~K and surface gravities of $3 < \logg < 4$~dex) with ages and RVs from LAMOST DR3 \citep{LAMOST} and astrometry from Gaia.
We took these raw data and performed linear regression in log(age)--log(\sigmavz) space to re-estimate the AVR for metal rich stars since they did not provide the intercept value for the formula.
We fit for $\alpha$ and $\beta$ in the relation: $\ln{age} = \beta \ln{(\sigma_{v_z} + \alpha)} $ and found $\alpha$ = -2.80 $\pm$ 0.53  and $\beta$ = 1.58 $\pm$ 0.19.
This value matches closely with the \cite{Yu2018} result.
We note that $\sim$ 20\% of the stars in our sample with LAMOST spectra have metallicities [Fe/H] $<-0.2$ dex.
This means we should use the AVR for metal-poor stars for these stars.
However, given that only a third of our sample have measured metallicities, it is not possible to fully account for this effect in our sample.
To maintain a homogeneous result, we only used the AVR for metal-rich stars, however we note that this is a drawback of this method and a future analysis with more complete measurements should account for metallicity.

Stars are binned in MTPR space because stars with similar rotation periods, temperatures, and absolute magnitudes are expected to have similar ages.
That stars with similar rotation periods and temperatures have similar ages is the underlying principle behind gyrochronology, so this is a reasonable assumption.
Binning in absolute magnitude allows stars of similar evolutionary states to be grouped together.
This is necessary because two stars with the same rotation period and temperature could have very different ages if one is on the main sequence and the other is a subgiant \citep{vansanders2013}.
An important caveat is that stars with similar absolute magnitudes, rotation periods and temperatures, but different metallicities, could have different ages.
Unfortunately, there are not enough stars in our sample to bin in metallicity too, although we acknowledge that variation in metallicity will likely add noise to our gyro-kinematic age estimates.
We also found that binning in Rossby number improved the accuracy of our results.
Although we are not entirely sure why this is, it may be because the relationship between effective temperature and Rossby number is more constant than the relationship between temperature and rotation period (for example, see Figure ~\ref{fig:parastars}).
Since our bins are rectangular, it is likely that transforming to Rossby number improves the age-binning.

We tested two different methods of binning stars: by 1) calculating the 4D phase-space distances and selecting the $N$ stars that are closest to the target star (nearest neighbor method), and by 2) constructing a 4D bin in phase-space with the target star in the center and selecting all the stars within a bin of a set size (binning method).

\begin{figure}[htp]
    \centering
    \includegraphics[width=0.47\textwidth]{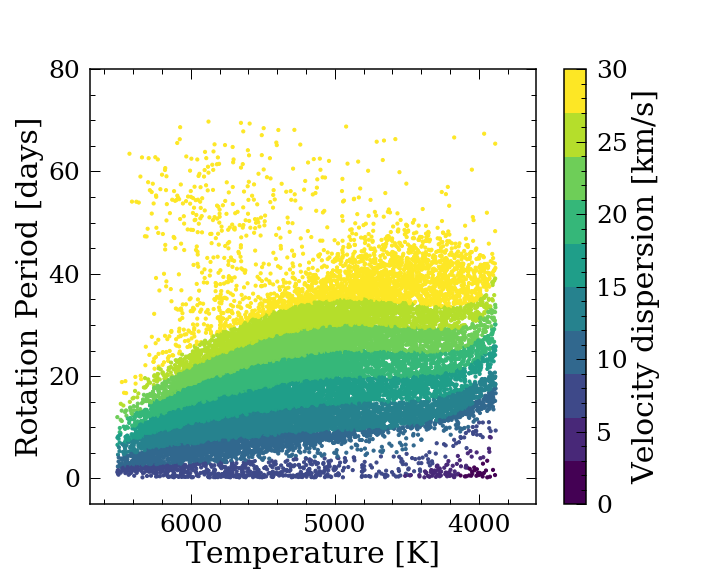}
    \caption{Simulated velocity dispersion for 29,949 stars with measured rotation periods based on the semi-empirical gyrochronology model described in \cite{Spada2020} and the AVR described in \citet{Yu2018}.}
    \label{fig:VDmock}
\end{figure}

To test these two methods, we generated mock vertical velocities with the semi-empirical gyrochronology model described in \cite{Spada2020} and tried to recover the result.
We first generated ages from the gyrochronology model, using the temperatures and rotation periods from the 29,949 stars with measured rotation periods.
We then converted these ages into \sigmavz using the AVR in \cite{Yu2018}.
We generated mock vertical velocities by drawing from a Gaussian with a mean velocity of 0~\kms\ and an uncertainty of \sigmavz.
Linear interpolation/extrapolation was used to calculate the ages of stars between/outside the \citet{Spada2020} model grid points. 
Figure~\ref{fig:VDmock} shows the simulated velocity dispersion in temperature versus rotation space. 

\begin{figure}htp
    \centering
    \includegraphics[width=0.47\textwidth]{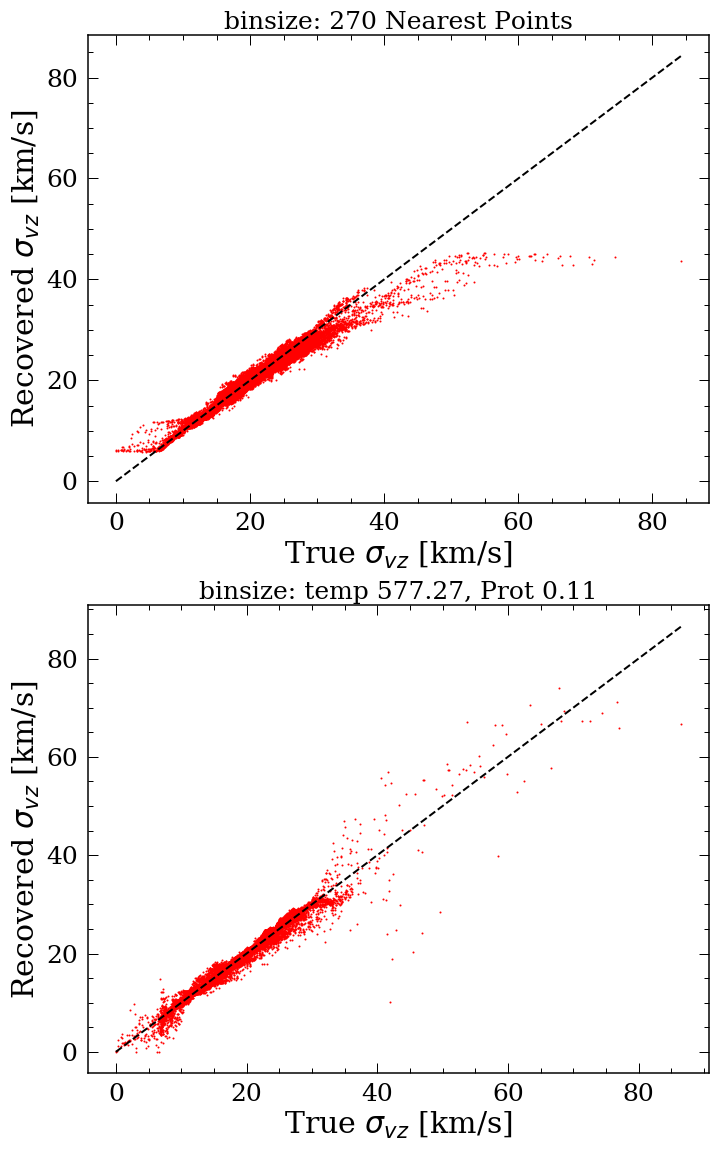}
    \caption{Recovery velocity dispersion of mocked data (\ref{fig:VDmock}) for using the binning method and the nearest stars method.
    The optimized number of nearest stars and bin size in \teff and \prot gave a minimum $\chi^2$ of 0.82 and 1.46, respectively.
    The nearest star method recovers poorly at the boundaries (when the velocity dispersion is extreme), and therefore, we decided to use the binning method to calculate gyro-kinematic ages.}
    \label{fig:mockResult}
\end{figure}

To establish which of these two methods provides the most accurate and precise gyro-kinematic ages, we simulated the velocity dispersions of stars and attempted to recover those dispersions.
Figure~\ref{fig:mockResult} shows the recovery results after optimizing the bin size and number of nearest neighbors by minimizing $\chi^2$.
The minimum reduced $\chi^2$ values were 0.82 for the binning method and 1.46 for the nearest stars method.
Figure \ref{fig:mockResult} shows the simulated velocity dispersions, compared with the measured velocity dispersions for the nearest neighbor and binning methods.
The top panel shows the results using the nearest neighbor method, and indicates that this method introduces bias in the recovered velocity dispersions: it over-predicts at low velocity dispersion and under-predicts at high velocity dispersion. 
This is caused by the boundaries: for stars at the edges of the period distribution, their nearest neighbors are all on one side.
As a result, the velocity dispersions of stars with the longest rotation periods will be calculated using stars that, on average, are younger than them and therefore have smaller velocity dispersions.
Similarly the velocity dispersions of stars with the shortest rotation periods are over-estimated because their nearest neighbors all have longer rotation periods and are older than they are.

The binning method does not experience bias at the boundaries.
However, the precision of the estimated velocity dispersion is poor in sparsely populated areas, where there are few stars in each bin.
Therefore, we also set a minimum and maximum number of stars for each bin, in which we decrease/increase the bin size by 10\% each time if the stars exceed the maximum/minimum number until the number of stars is between the minimum and maximum.
The lower limit ensures there are enough stars in each bin so the velocity dispersion calculated is relatively precise, and the upper limit ensures that we are not including stars with a large range of ages.
Because it is less biased, we decided to use the binning method to estimate gyro-kinematic ages for these stars in MTPR space.
The gyro-kinematic ages we infer with this method depends on the choice of bin size, so we optimized the bin size using a sample of measured stellar ages.

\subsubsection{Optimizing the Bin Size}\label{subsub:binsize}
In this section, we describe the process of optimizing the bin size in MTPR space, as well as the minimum and maximum numbers of stars in each bin in order to achieve the best agreement with the stars with known age measurements from asteroseismology and isochrone fitting.

We first performed $\chi^2$ tests with respect to stars with asteroseismic ages from the Kepler LEGACY sample \citep{Silva2017}. This LEGACY sample includes 66 main-sequence stars analyzed with seven different stellar models.
The uncertainties for these asteroseismic ages were calculated by taking the range of the results from these seven different models.
After cross-matching the asteroseismic stars with our star sample, we found 13 stars in common.
We then excluded stars that appeared to be experiencing weakened magnetic braking \citep{vansaders2016}.
At $Ro\approx$2, the magnetic braking mechanism of stars is thought to lose efficiency and, as a result, stars stop spinning down.
Although our gyro-kinematic dating method is not technically incompatible with stars that are no longer spinning down, it can only calculate accurate ages for stars that are well-represented in our data set, and it is not clear how many stars in the \citet{McQuillan2014} catalog have reached this critical Rossby number and are experiencing weakened braking \citep{Saders2019}.
Weakened-braking stars may rotate more rapidly at the same age and temperature than non-weakened-braking stars.
For this reason, the gyro-kinematic ages calculated for these stars will be underestimated from the inclusion of younger stars in the reference set, and vice versa.
To avoid this bias, we removed these stars from the sample.
To identify weakened-braking stars, we predicted the rotation periods for the asteroseismic stars using the non-weakened gyrochronology model in the \texttt{stardate} software package \citep{Angus2019a, Angus2019b}.
We excluded seven stars with predicted rotation periods differing by more than five days from their measured rotation periods, leaving us with six stars.
Although it is beyond the scope of this paper to investigate weakened magnetic braking from a kinematic age standpoint, we note that a preliminary examination indicates that the asteroseismic stars removed from our sample show good age agreement with the gyro-kinematic ages of stars in table 2 of \citet{McQuillan2014}, which include stars with low measurement confidence that could be going through magnetic braking.

To find the optimum bin size, we performed a 10 $\times$ 10 $\times$ 10 $\times$ 10 grid search with temperatures between 50 K and 500 K, and $M_G$, log$_{10}$(rotation period), and log$_{10}$(\ro) between 0.1 dex and 0.5 dex.
We compared gyro-kinematic ages with the asteroseismic ages to calculate $\chi^2$ at each point in this grid.
We took this approach to keep computation time low.
This yielded a bin size of $M_G$ = 0.46 mag, log$_{10}$(period) = 0.14 dex, log$_{10}$(\ro) = 0.14 and temperature = 150 K with a $\chi^2$ of 0.05.
We then estimated the uncertainty on these gyro-kinematic ages by performing bootstrapping on the parameters we used ($M_G$, rotation periods, \ro, temperature, and vertical velocity) with 50 random samplings and computed the standard deviation of the resulting gyro-kinematic ages as the uncertainty.
In order to make sure 50 sampling is enough, we also estimated the uncertainties using 100 random samplings for ten random stars and the uncertainty did not change.

Next, we performed similar $\chi^2$ tests for Kepler stars with isochrone ages. 
\cite{Berger2020} produced a catalog of stellar properties for 186,301 \kepler stars based on Gaia data, 
with median isochrone age uncertainties of 2.71 Gyr (54\%).
To optimize the bin size with respect to these isochronal ages, we first excluded stars older than 10 Gyr, where there is an overdensity that is likely due to systematic errors in the model.
Stars with age uncertainties greater than 3 Gyr were also excluded, which removes most of the K and M dwarfs from our comparison sample.
After excluding these stars, we were left with 8,958 stars with isochrone ages with measured rotation periods to compare gyro-kinematic ages with, in which most of these stars are subgiants and F/G dwarfs.

We then minimized the $\chi^2$ by performing a similar grid search as done for the asteroseismic stars; we also changed the minimum and maximum number of stars allocated to each bin.
The large sample size of isochrone ages allow us to determine the requisite minimum and maximum number of stars in each bin.
As mentioned in the previous section, the lower limit for the number of stars in each bin ensures there are enough stars to calculate precise velocity dispersion, and the upper limit ensures that we are not including stars with a large range of ages.
The number density of the stars in the MTPR space indicates the rate at which these stars move in the phase space, in which a higher density region corresponds to a slower evolution rate and vice versa.
As a result, setting a fixed number of stars in each bin ensures that stars in each bin are of a similar age.
In other words, the bin size grows in regions of parameter space where isochrones are spaced farther apart (e.g., on the subgiant branch) and shrinks where isochrones are tightly packed (e.g., on the main sequence).
If the number of stars exceeded the minimum or maximum limit, the bin was reduced/increased by 10\% until the number of stars in each bin was within the limits or the bin size had changed by more than 100\% from its original bin size.
The minimum and maximum number of stars in each bin were optimized to be 15 and 30, respectively.

The optimized bin size for the isochrone ages was $M_G$ = 0.5 mag, log(\prot) = 0.1 dex, \ro = 0.4, and \teff\ = 500 K, with a reduced $\chi^2$ of 0.8; the bin sizes for $M_G$ and log(\prot) are on par with those found for the asteroseismic stars, whereas the \ro and \teff\ bins are much larger.
These large bin sizes are concerning, so we also calculated $\chi^2$ for the isochrone ages using the optimized bin size from the asteroseismic stars and obtained a $\chi^2$ value of 1.02.
Since this $\chi^2$ value was similar to the {\it minimized} $\chi^2$ for the isochrone ages, we decided to use the bin size from the asteroseismic stars to determine all the gyro-kinematic ages since the asteroseismic ages are more precise.

We then estimated the uncertainties on the gyro-kinematic ages with the bootstrapping method described in the previous paragraphs but with the optimized bin size.
This yielded a median age uncertainty from bootstrapping to be 1.30 Gyr across all ages.

We also added the uncertainties from the AVR fit. 
In which we calculated the uncertainty, using error propagation, to be $\sigma_{\ln{age}}=\sqrt{(\ln{\sigma_{v_z}}\sigma_\beta)^2+(\sigma_\alpha)^2+((\beta/\sigma_{v_z})\sigma_{\sigma_{v_z}})^2}$, in which $\sigma_\alpha$, $\sigma_\beta$, and $\sigma_{\sigma_{v_z}}$ are uncertainties on $\alpha$, $\beta$, and $\sigma_{v_z}$, respectively. The uncertainty for the vertical velocity dispersion is the bootstrapping uncertainties.
This yields a median uncertainty from the AVR to be 1.76 Gyr across all ages.

\begin{figure}[htp]
    \centering
    \includegraphics[width=0.5\textwidth]{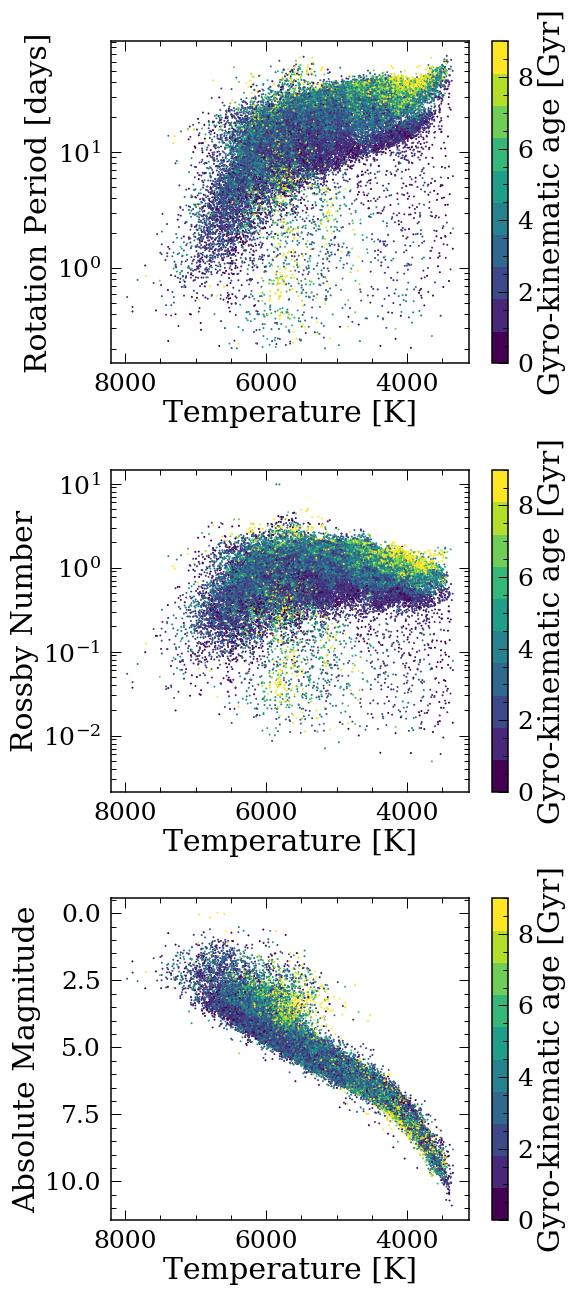}
    \caption{2D projections of 29,949 stars in MTPR space colored by gyro-kinematic ages. The rotation period/\ro\ slows down as a star ages and the evolved and low-mass stars are the oldest stars in our catalog.}
    \label{fig:parastars}
\end{figure}

Figure~\ref{fig:parastars} shows the 2D projections of the stars in MTPR space colored by their gyro-kinematic ages.
The rotation period and \ro versus $T_\mathrm{eff}$ space (top and middle panels) shows that stellar rotation slows down as stars age, as expected.
Converting rotation period to Rossby number flattens the upward trend seen in the coolest stars in the top panel.
This conversion effectively transforms the data to a shape that is more compatible with the square-shaped bins we use to calculated gyro-kinematic ages.
Including Rossby number as an additional dimension in the binning process improves age accuracy by $\sim$ 10\%.
The old (yellow) stars with $5000 < \teff < 6000$~K seen in all three panels are subgiants.
These stars exhibit a range of rotation periods as they expand off of the main sequence, as predicted by theoretical models \citep{vansanders2013, Saders2019}.

\section{Results}\label{sec:results}
\subsection{Comparing Gyro-Kinematic Ages with Benchmark Ages}\label{subsub:comp}
After optimizing the bin size for gyro-kinematic procedure, we calculated gyro-kinematic ages for members of stars clusters and for two white dwarf--M dwarf binaries.
For the former, we used the \cite{Curtis2020} catalog which is an assembly of rotation periods from a variety of benchmark clusters, including the Pleiades \citep[{120 Myr;}][]{Rebull2016}, Praesepe \citep[{670 Myr;}][]{Douglas2017,Douglas2019}, Hyades \citep[{730 Myr;}][]{Douglas2016,Douglas2019}, NGC 6811 \citep[{1 Gyr;}][]{Meibom2011,Curtis2019}, NGC 752 \citep[{1.4 Gyr;}][]{Agueros2018}, NGC 6819 \citep[2.5 Gyr;][]{Meibom2015}, and Ruprecht 147 \citep[{2.7 Gyr;}][]{Curtis2020}. This catalog provides Gaia DR2 data and classifies subsets of each cluster as ``benchmark rotators'' if they satisfy criteria for single-star membership outlined in Appendix A.4 of that paper. 

Kiman et al. in prep. studied the ages of low-mass stars using white dwarf--M dwarf binary systems. 
They developed \texttt{wdwarfdate},\footnote{https://wdwarfdate.readthedocs.io/en/latest} an open source Python code which estimates the age of a white dwarf from an effective temperature and a surface gravity in a Bayesian framework.
The white dwarf age is the sum of its cooling time and the main sequence lifetime of its progenitor.
The cooling time is obtained from the spectroscopic properties using cooling models \citep{Bergeron_1995}.\footnote{\url{http://www.astro.umontreal.ca/~bergeron/CoolingModels/}} 
To determine the main sequence lifetime of the progenitor star, the progenitor's mass is estimated from the WD mass using an initial-to-final mass relation \citep[IFMR;][]{Cummings2018}, and then the progenitor lifetime is adopted from a MESA isochrone \citep[MIST;][]{Dotter2016}. We found two \kepler stars that have measured rotation periods and are comoving with a white dwarf (KIC 11075611 and KIC 12456401).
We age-dated the white dwarfs with \texttt{wdwarfdate}, using the spectroscopic measurements for effective temperature and surface gravity from \cite{gaiawd}. 
We do not have a large sample of these stars but these binaries will be more powerful in the future when a larger sample is available. 

In order to obtain the gyro-kinematic ages for stars in clusters, for each target star in the cluster, we calculated the velocity dispersion of the target star using the optimal bin size determined in the previous section at the $T_\mathrm{eff}$, $M_G$, \prot, and \ro for the target star as the bin center.
We then calculated the age using the \cite{Yu2018} AVR.
The gyro-kinematic cluster ages and uncertainties were then calculated by taking the mean and standard deviation of the gyro-kinematic ages of all the stars in the cluster.

\begin{figure*}[htp]
    \centering
    \includegraphics[width=\textwidth]{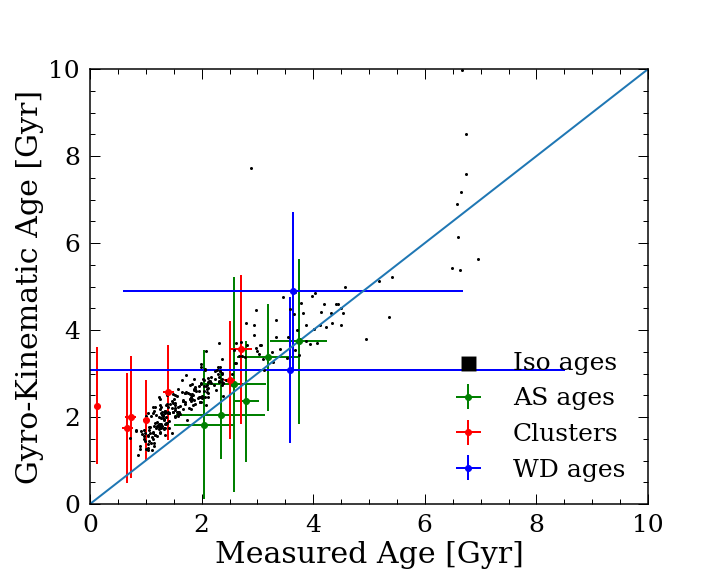}
    \caption{Comparison of gyro-kinematic ages with literature ages for benchmarks, including 
    stars with isochrone ages \citep[Iso;][]{Berger2020}, 
    Kepler LEGACY stars characterized with asteroseismic ages \citep[AS;][]{Silva2017}, 
    members of star clusters \citep{Curtis2020}, and 
    white dwarf--M dwarf binaries \citep[WD;][]{Kiman2020}.
    We overestimate ages of very young cluster likely due to the presence of older tight binaries contaminating the reference bins used to calculate the velocity dispersion used to assign their gyro-kinematic ages.
    The black dots show the results for the isochrone stars, where we have applied extreme deconvolution to aid the visual interpretation.
    The systematic offset between isochrone ages and gyro-kinematic ages could be due to bias in the binning method and/or an offset between the age scales for the MIST isochrone models \citep{Choi2016} used in \cite{Berger2020} and our adopted AVR.
    RMS = 1.22 Gyr, MAD = 1.08 Gyr for the clusters, and RMS = 0.26 Gyr, MAD = 0.20 Gyr for asteroseismic stars.
    Another version of this graph without applying extreme deconvolution on the isochrone ages comparison is shown in figure~\ref{fig:protkin_noXD} in the Appendix section. }
    \label{fig:protkin}
\end{figure*}

Figure~\ref{fig:protkin} compares the gyro-kinematic ages for these benchmarks with their literature ages.
In order to better compare gyro-kinematic ages, which are average ages, with isochrone ages, which are individual ages, we also took the average of the isochrone ages with the same optimized bin in MTPR space to get rid of outliers and applied extreme deconvolution (\texttt{XD}), which is useful for disentangling underlying distributions from noisy data.
This method finds the underlying density distribution from noisy, heterogeneous and incomplete data using a Gaussian mixture model approach \citep{Bovy2011}. 
Without applying extreme deconvolution, the high level of scatter caused by large age uncertainties prevents a meaningful visual comparison between gyro-kinematic and isochrone ages.
In other words, the comparison plot looks like a scatter plot with little correlation.
Extreme deconvolution predicts the noise-free data and provides a guide for the eye, allowing a more meaningful comparison to be drawn.
The black dots in Figure \ref{fig:protkin} show 500 samples from the Gaussian that best describes the underlying noise-free distribution, estimated using extreme deconvolution, for comparison against isochrone ages.
The root mean square (RMS) and median absolute deviation (MAD) between the gyro-kinematic ages and benchmark ages are RMS = 1.22 Gyr, MAD = 1.08 Gyr for the clusters, and RMS = 0.26 Gyr, MAD = 0.20 Gyr for asteroseismic stars.

In general, the gyro-kinematic ages agree well with other benchmark ages.
There are two obvious disagreements---a deviation between gyro-kinematic ages and ages of very young clusters ($<$ 1 Gyr), and a systematic offset between gyro-kinematic ages and isochrone ages.
One reason why we may overestimate ages for the young clusters is that stars within these clusters generally spin rapidly. 
There is mounting evidence suggesting that around 60\% of G to early-M stars that rotate faster than 7-10 days are synchronized binaries \citep[e.g.,][]{simonian2019, Angus2020_kinage}.
Such rapidly rotating stars in binary with short orbital periods, which tend to be old, can be wrongly assigned to a bin with rapidly rotating young stars.
Therefore, averaging over a bin of stars with short rotation periods will most likely result in over-estimating the age for the targeted stars.

The systematic offset between isochrone ages and gyro-kinematic ages is most likely caused by a bias in either the isochrone model or the AVR.
However, the offset between the isochrone ages and the gyro-kinematic is only around 1 Gyr at maximum, which is typical when comparing ages estimated with different models \citep{Silva2017}.

\section{Discussion \& Future Work}\label{sec:disc}
\subsection{Advantages \& Limitations}\label{subsec:limit}
Using gyro-kinematic ages comes with several caveats:
\begin{itemize}
    \item {\it Average ages}---
    gyro-kinematic ages do not necessarily reflect the actual age of each individual star.
    The gyro-kinematic age of each star is assigned according to the kinematics (i.e., vertical velocity dispersion) of stars with similar rotation periods, temperatures, Rossby numbers, and luminosities. This means that, by design, gyro-kinematic ages vary smoothly across MTPR space, with a characteristic length-scale set by the bin size.
    This means the gyro-kinematic age for a star that has abnormal stellar properties or has followed an atypical path of stellar evolution (e.g., a blue straggler or tidally-synchronized binary) will most likely be incorrect.
    For example, a star that spins anomalously rapidly compared to analogous stars of a similar age will have an under-estimated age from gyro-kinematics.
    Although this caveat should be acknowledged when using gyro-kinematic ages, this feature is shared by all age-dating techniques. 
    Relations that are calibrated using data with some amount of intrinsic scatter (e.g., caused by stars that still exhibit their initial conditions) that is not captured by measurement uncertainties (as is true of all stellar rotation data) will only reflect the average behaviour of stars.
    \item {\it Uncertainties}---It is hard to estimate the true uncertainties for gyro-kinematic ages. 
    We have taken into account the formal uncertainties on the stellar parameters (temperature, absolute magnitude, rotation period, Rossby number, and vertical velocity). 
    We tested how the bin size might affect the gyro-kinematic ages by perturbing the bin size in each stellar parameter space 10 times by up to 20\% and calculated the standard deviation of the ages.
    For example, the optimized bin size is [$M_G$, period, \ro, temperature] = [0.46 dex, 0.14 dex, 0.14, 150 K], and we tested how much the temperature bin size might have affected the results by keeping the bin size in other stellar parameters unchanged and only changing the temperature bin size between 120 K and 180 K with ten equally spaced values.
    However, it is hard to estimate the error from changing the bin size since we do not know exactly how much perturbation around the optimal bin size is needed to obtain the appropriate uncertainty.
    As a result, we did not include the uncertainty from the effect of changing the bin size, however varying the bin sizes by 20\% caused the gyro-kinematic ages to fluctuate by 10\%.
    \item {\it Ages for rapid rotators and old G stars}---As pointed out previously (e.g., Section \ref{sec:results}), we were not able to estimate accurate ages for young rapidly rotating stars most likely because tight binaries are contaminating their MTPR bins.
    In Sections \ref{sec:intro} and \ref{sec:datamethod}, we also pointed out that we would not be able to estimate ages for stars with Rossby numbers greater than $\gtrsim$2.
    Since these stars are either going through weakening braking \citep{vansaders2016} or reaching the period detection limit for inactive stars or both, they may not be well represented in the \citet{McQuillan2014} rotation sample.

\end{itemize}

However, despite these limitations, the gyro-kinematic ages and other benchmark ages are mostly in agreement, and gyro-kinematic ages do have some advantages over other age-dating methods.
A major advantage is
that we are able to estimate ages for low-mass stars.
Furthermore, we only made two simple assumptions to obtain gyro-kinematic ages---that stars in the same bin in stellar parameter space (in this case, \prot, \teff, \ro, and $M_G$) are similar in age and that vertical velocity dispersion increases over time.
Both of these assumptions have been tested extensively by gyrochronology studies \citep[e.g.,][]{Barnes2003, Curtis2020} and kinematic studies \citep[e.g.,][]{Stromberg1946,Yu2018}.
The only model we included was the AVR, and as a result the absolute ages of these stars are subject to the accuracy of that \citet{Yu2018} AVR, but the relative age ranking should be correct.

\subsection{Gyro-Kinematic Ages from Predicted Periods}\label{subsec:abund}
Most stars observed by Kepler do not have a measured rotation period. 
However, it may still be possible to calculate their gyro-kinematic ages by {\it predicting} their rotation periods using our machine learning method called \texttt{Astraea} \citep{Lu2020}\footnote{Avaliable at \url{https://github.com/lyx12311/Astraea}}.
Using Random Forest, this method predicts the rotation periods of stars observed by \kepler based on \teff, \logg, luminosity, \gaia photometry and its error, velocities, radius, galactic latitude, and photometric variability (\rvar).
This is important because only one in five \kepler stars has a period measured by \citet{McQuillan2014} with confidence.
The remaining 80\% of stars did not have detectable or measurable rotation periods for a few potential reasons: their variability was low-amplitude, their rotation periods were very long, their variability was quasi- or aperiodic, or they were extremely faint.
In \citet{Lu2020} we demonstrated that {\tt Astraea} can precisely predict the periods of stars with measured rotation periods, however we did not establish whether {\tt Astraea} can predict the rotation periods of stars {\it without} measured rotation periods.
Can we use our machine learning tool to predict the rotation periods of stars that do not have rotation periods measured with high confidence?

\begin{figure}[htp]
    \centering
    \includegraphics[width=0.5\textwidth]{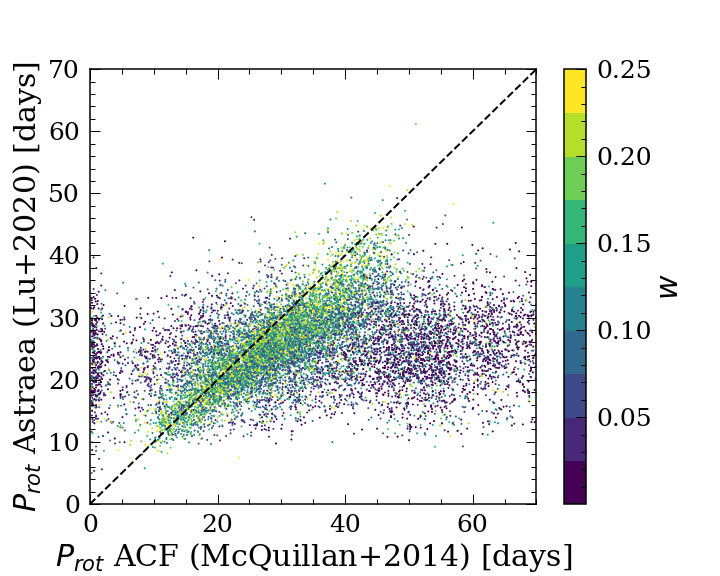}
    \caption{Predicted rotation periods versus measured rotation periods from \cite{McQuillan2014} Table~2 (i.e., those with no significant period detection), color-coded by the measurement confidence weight ``$w$'' from their paper.
    We are able to predict rotation periods with high ``$w$'' values ($>$0.15).
    This means, we are able to increase the size of the validated rotation periods described in \cite{McQuillan2014} validated by up to 25\%}
    \label{fig:McQTb2}
\end{figure}

One way to test this is to predict the rotation periods of stars that were not used to train the random forest algorithm, but that do have measured rotation period measurements.
\citet{McQuillan2014} included a table of stars for which the rotation periods have low confidence levels, parameterized as ``$w$'' for weight (see their table 2).
These stars were not used to train {\tt Astraea}, however we find that we can accurately predict many of their rotation periods.
Figure~\ref{fig:McQTb2} shows the prediction results for stars with low $w$ rotation periods from that table. 
The weights were assigned according to the height of the primary autocorrelation function (ACF) peak with respect to the troughs on each side (local peak height; LPH) and position in \teff--LPH--period space.
As shown in the figure, most of the stars with high weights (green or yellow colored points) lie on the equality line and the stars whose measured and predicted rotation periods do not agree have low weights (purple colored points).
This suggests \texttt{Astraea} is able to predict accurate rotation periods in cases where the signal amplitude is above some threshold.
In cases where stars have low weights, the periods predicted with \texttt{Astraea} disagree with those measured by \citet{McQuillan2014}.
This could be because either one or both methods are incorrect, and we are not able to tell whether we are able to predict accurate rotation periods for these stars.
Regardless of the reason, we suggest using a new weight, 0.15, for selecting validated rotation periods in the catalog described in \cite{McQuillan2014}.
By doing so, we are able to increase the catalog size up to 25\%.

Another way to test the predicted periods is to see whether the velocity dispersion of stars increases with predicted rotation period, $M_G$ and temperature in a way that is similar to the trends observed in the sample of stars with directly measured rotation periods.
We predicted the rotation periods of 45,000 stars observed by Kepler, but without reported rotation periods.
These stars were selected after applying cuts to remove subgiants and photometric binaries, following the procedure described in \citet{Lu2020}.
We also excluded stars with stellar parameters outside of the \kepler training set described in \cite{Lu2020}.
Only predicting periods for stars within the range of the stellar parameters of the training set is extremely important since most machine learning algorithms, including Random Forests used by \texttt{Astraea}, cannot extrapolate outside of the training data.
Rotation periods were then predicted with \texttt{Astraea} using the absolute magnitudes, temperatures, radii, colors, kinematics, and the light curve variability of these stars.
Gyro-kinematic ages were calculated with the same method described in section \ref{subsec:method}, using these predicted rotation periods.

We find tentative evidence to suggest that many of the rotation periods predicted  by {\tt Astraea} are accurate, or at least correctly ranked.
For example, we find that the {\tt Astraea}-predicted periods are correlated with \citet{Berger2020} isochrone ages.
As shown in the following section, we also find that gyro-kinematic ages calculated from predicted periods can reproduce some of the age-abundance trends previously measured in a sample of Solar twins \citep{Bedell2018}.
However, although this technique shows promise, it still requires some refinement and optimization.
A thorough quantification of the precision and accuracy of gyro-kinematic ages, calculated with {\it predicted} rotation periods, is beyond the scope of this paper and we leave that to a future exercise.

\subsection{Detailed age-element abundance trends with gyro-kinematic ages from predicted periods}
To explore the validity of our gyro-kinematic ages, calculated using predicted rotation periods, we investigated the age-abundance trends that have been found for Solar twins.
Detailed age-chemical abundance trends measure the time-dependent chemical evolution of the gas in the Milky Way disk.
These trends offer strong empirical constraints on nucleosynthesis processes, as well as the mixing of star forming gas in the disk.
We wanted to compare our results with those found by \cite{Bedell2018} to see whether we can recover the same age-abundance trends using gyro-kinematic ages.
If so, gyro-kinematic ages could provide a new avenue for Galactic chemical evolution studies.
To provide a meaningful comparison to the \cite{Bedell2018} results, we focused on Solar twins.

We cross-matched our catalog with APOGEE DR16 \citep{DR162020,Majewski2017} using the same criteria described in \cite{Bedell2018} to select our Solar twin samples (\logg\ within 0.1 dex, [Fe/H] within 0.3 dex, and temperature within 100 K of the Sun), and found 108 stars. 
We then calculated gyro-kinematic ages using \texttt{Astraea}-predicted periods to further validate our method and to test whether gyro-kinematics could be useful for future age--abundance studies.
We used {\it predicted} rather than measured rotation periods as most stars in our Solar twin sample did not have measured rotation periods.
Only 12 stars with very similar ages (1-3 Gyr) have gyro-kinematic ages from measured periods, and such a narrow age range means these stars were not able to provide constrained age--element abundance trends.

\begin{figure*}[htp]
    \centering
    \includegraphics[width=\textwidth]{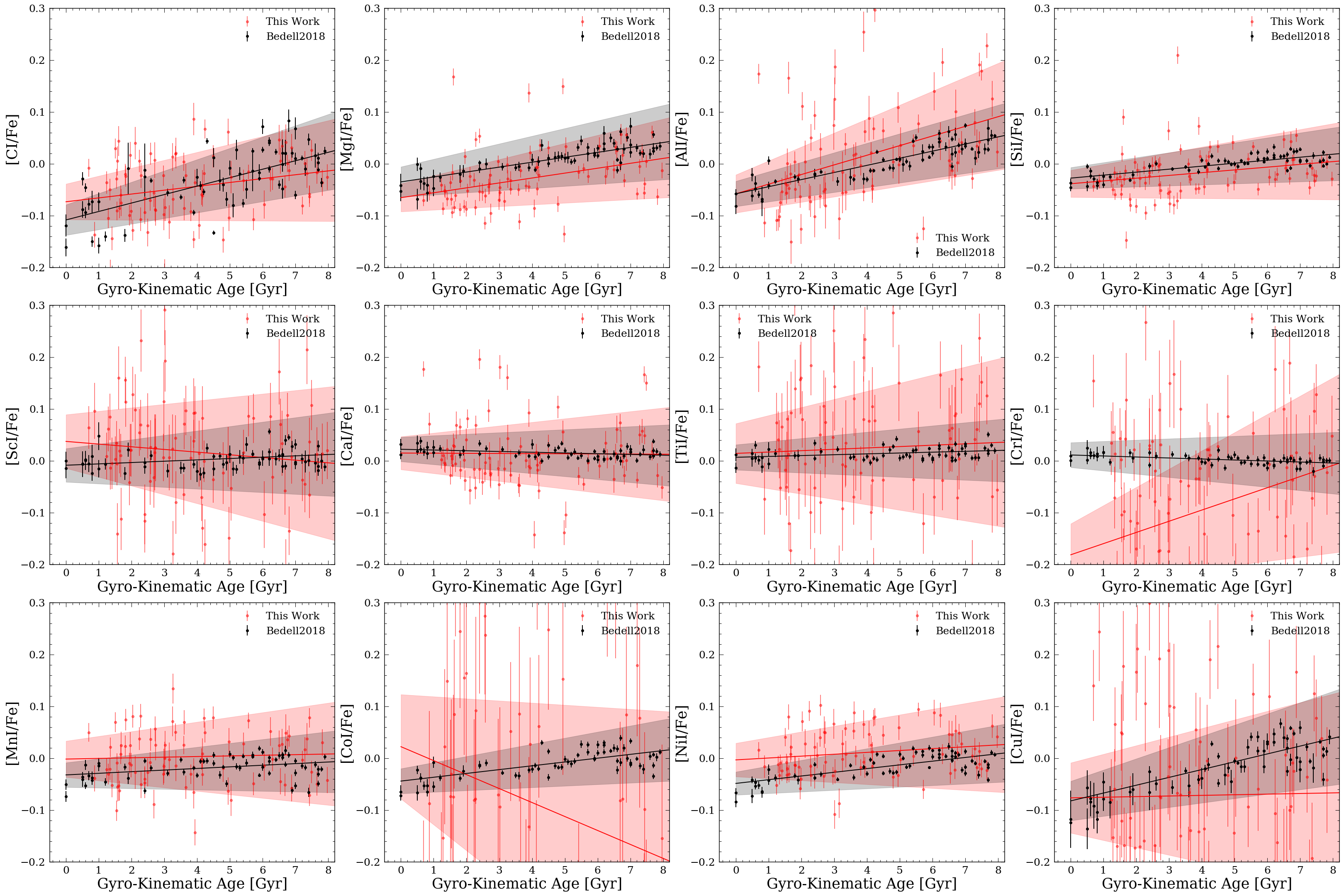}
    \caption{Individual age--element abundance trend with gyro-kinematic ages for 108 solar twins. 
    The black dots and black lines are the results from \cite{Bedell2018}, and the red dots and red lines are results from this work. 
    The shaded area represents the 1-$\sigma$ uncertainty from the fit, estimated using the covariance matrix.
    In general, the abundance--age trends agree with those from \cite{Bedell2018} except for ones with large scatter; e.g., Cr, Co, and Cu}
    \label{fig:Abundance}
\end{figure*}

To measure the trends, we excluded any stars with a gyro-kinematic upper age limit greater than 8 Gyr, as these stars could be stars from the high-$\alpha$ disk and exhibit a different trend to those from the low-$\alpha$ disk \citep{Bedell2018}.
Figure~\ref{fig:Abundance} shows the comparison between this work and 79 solar twins from \cite{Bedell2018}.
We calculated the age--abundance trends by performing linear fits using \texttt{numpy.polyfit}, and the uncertainties were estimated from the covariance matrix. 
In general, our trends are very similar with those from \cite{Bedell2018} within the uncertainties.
The ones that do not agree (Cr, Co, and Cu) have large uncertainties on their abundances measurements from APOGEE.

The study of age--abundance trends using gyro-kinematic ages from predicted periods shows promising results.
It may be possible to analyze individual age--abundance trends by combining gyro-kinematic ages with APOGEE abundances for a larger sample of stars in the future.

\section{Conclusion}\label{sec:concl}
In this paper, we created a gyro-kinematic catalog containing 29,949 stars with measured rotation periods
from \citet{McQuillan2014}, \citet{Garcia2014}, and \citet{Santos2019}.
We obtained the ages by binning the stars in MTPR space with an optimized bin size of [$M_G$, period, \ro, temperature] = [0.46 dex, 0.14 dex, 0.14, 150 K]. These bin sizes produced the smallest average $\chi^2$ value with respect to the LEGACY asteroseismic stars from \citet{Silva2017}.
We also found the optimum minimum and maximum number of stars in each bin to be 15 and 30 stars, respectively, by minimizing the average $\chi^2$ respect to isochrone ages from \cite{Berger2020}. 
We optimized the min/max number of stars with respect to isochrone ages since we did not have enough stars with asteroseismic ages to constrain these two parameters.
If the number of stars exceeded the minimum or maximum limit, the bin was reduced/increased by 10\% until the number of stars in each bin was within the limit or the bin size had changed more than 100\% from its original bin size.
We estimated the uncertainties for these ages by perturbing the stellar parameters ($M_G$, rotation period, \ro, temperature, and vertical velocity) within their errors via bootstrapping with a sample size of 50.

We compared the gyro-kinematic ages with other benchmark ages (isochrone ages from \citealt{Berger2020}, asteroseismic ages from \citealt{Silva2017}, cluster ages from \citealt{Curtis2020}, and WD--M dwarf binary ages from \citealt{Kiman2020}).
Aside from the offset between ages from this work and isochrone ages, gyro-kinematic ages agree well with other benchmark ages.
We calculated the RMS and MAD between the gyro-kinematic ages and the benchmark ages and obtained a RMS = 1.22 Gyr, MAD = 1.08 Gyr for the clusters, and RMS= 0.26 Gyr, MAD = 0.20 Gyr for asteroseismic stars.
The offset between the isochrone ages and gyro-kinematic ages could be caused by systematic errors in the stellar evolution models used to calculate the isochrone ages or the AVR we used to calculate gyro-kinematic ages.

In section \ref{sec:disc}, we also estimated gyro-kinematic ages for \kepler stars with rotation periods predicted from \texttt{Astraea}.
We did this so we could test whether or not the Random Forest method described in \cite{Lu2020} is able to accurately predict periods.
We first predicted rotation periods for stars included in table 2 of \cite{McQuillan2014}, and we found we were able to predict stars with weight, $w >$ 0.15.
We suggested a new $w$ cut of 0.15 to select for stars with measured rotation periods and by doing so, increased their catalog size by up to 25\%.
After predicting the periods, we estimated the ages with the same method used for stars with measured rotation periods, we cross-matched our sample with APOGEE DR16 and found 108 stars with detailed abundances measured.
We plotted the detailed age-element abundance relation for 12 elements and showed that our result mostly agree with those from \cite{Bedell2018}.
This result suggest getting gyro-kinematic ages from predicted periods could be promising.

\acknowledgments
This work made use of the \url{gaia-kepler.fun} cross-match database created by Megan Bedell.
This paper includes data collected by the \kepler mission. Funding for the \kepler mission is provided by the NASA Science Mission directorate. 

This work has made use of data from the European Space Agency (ESA) mission
\gaia (\url{https://www.cosmos.esa.int/gaia}), processed by the \gaia
Data Processing and Analysis Consortium (DPAC,
\url{https://www.cosmos.esa.int/web/gaia/dpac/consortium}). Funding for the DPAC
has been provided by national institutions, in particular the institutions
participating in the {\it Gaia} Multilateral Agreement.

R.A. acknowledges support from NASA award 80NSSC20K1006.

Guoshoujing Telescope (the Large Sky Area Multi-Object Fiber Spectroscopic Telescope LAMOST) is a National Major Scientific Project built by the Chinese Academy of Sciences. Funding for the project has been provided by the National Development and Reform Commission. LAMOST is operated and managed by the National Astronomical Observatories, Chinese Academy of Sciences.

This research has also made use of NASA's Astrophysics Data System, 
and the VizieR \citep{vizier} and SIMBAD \citep{simbad} databases, 
operated at CDS, Strasbourg, France.

%

\vspace{5mm}
\facilities{\textit{Gaia}, \textit{Kepler}, LAMOST, APOGEE}


\software{Astropy \citep{astropy:2013, astropy:2018}, Numpy \citep{oliphant2006guide}, Astraea \citep{Lu2020}, Pandas \citep{reback2020pandas,mckinney-proc-scipy-2010}, Matplotlib \citep{hunter2007matplotlib}, Jupyter Notebook \citep{Kluyver:2016aa}, wdwarfdate \citep{Kiman2020}}

\appendix \label{sec:append}
\renewcommand{\thesubsection}{\Alph{subsection}}
\counterwithin{figure}{subsection}
\counterwithin{table}{subsection}
\subsection{Additional figures}
\begin{figure*}[htp]
    \centering
    \includegraphics[width=\textwidth]{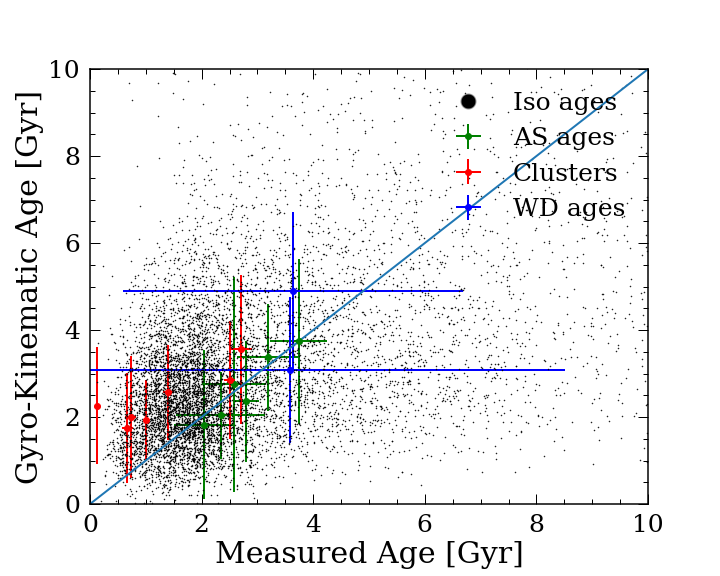}
    \caption{Same as figure~\ref{fig:protkin} but showing the comparison between gyro-kinematic ages and isochrone ages from \cite{Berger2020} without extreme deconvolution.}
    \label{fig:protkin_noXD}
\end{figure*}

Figure~\ref{fig:protkin_noXD} shows the same comparison between benchmark ages and gyro-kinematic ages without applying extreme deconvolution to the isochrone ages comparison.



\bibliography{references.bib}{}

\begin{thebibliography}{}
\expandafter\ifx\csname natexlab\endcsname\relax\def\natexlab#1{#1}\fi
\providecommand{\url}[1]{\href{#1}{#1}}
\providecommand{\dodoi}[1]{doi:~\href{http://doi.org/#1}{\nolinkurl{#1}}}
\providecommand{\doeprint}[1]{\href{http://ascl.net/#1}{\nolinkurl{http://ascl.net/#1}}}
\providecommand{\doarXiv}[1]{\href{https://arxiv.org/abs/#1}{\nolinkurl{https://arxiv.org/abs/#1}}}

\bibitem[{{Ag{\"u}eros} {et~al.}(2018){Ag{\"u}eros}, {Bowsher}, {Bochanski},
  {Cargile}, {Covey}, {Douglas}, {Kraus}, {Kundert}, {Law}, {Ahmadi}, \&
  {Arce}}]{Agueros2018}
{Ag{\"u}eros}, M.~A., {Bowsher}, E.~C., {Bochanski}, J.~J., {et~al.} 2018,
  \apj, 862, 33, \dodoi{10.3847/1538-4357/aac6ed}

\bibitem[{{Ahumada} {et~al.}(2020){Ahumada}, {Allende Prieto}, {Almeida},
  {Anders}, {Anderson}, {Andrews}, {Anguiano}, {Arcodia}, {Armengaud},
  {Aubert}, {Avila}, {Avila-Reese}, {Badenes}, {Balland }, {Barger},
  {Barrera-Ballesteros}, {Basu}, {Bautista}, {Beaton}, {Beers}, {Benavides},
  {Bender}, {Bernardi}, {Bershady}, {Beutler}, {Bidin}, {Bird}, {Bizyaev},
  {Blanc}, {Blanton}, {Boquien}, {Borissova}, {Bovy}, {Brand t}, {Brinkmann},
  {Brownstein}, {Bundy}, {Bureau}, {Burgasser}, {Burtin}, {Cano-D{\'\i}az},
  {Capasso}, {Cappellari}, {Carrera}, {Chabanier}, {Chaplin}, {Chapman},
  {Cherinka}, {Chiappini}, {Doohyun Choi}, {Chojnowski}, {Chung}, {Clerc},
  {Coffey}, {Comerford}, {Comparat}, {da Costa}, {Cousinou}, {Covey}, {Crane},
  {Cunha}, {da Silva Ilha}, {Dai}, {Damsted}, {Darling}, {Davidson}, {Davies},
  {Dawson}, {De}, {de la Macorra}, {De Lee}, {de Andrade Queiroz}, {Deconto
  Machado}, {de la Torre}, {Dell'Agli}, {du Mas des Bourboux},
  {Diamond-Stanic}, {Dillon}, {Donor}, {Drory}, {Duckworth}, {Dwelly},
  {Ebelke}, {Eftekharzadeh}, {Eigenbrot}, {Elsworth}, {Eracleous},
  {Erfanianfar}, {Escoffier}, {Fan}, {Farr}, {Fern{\'a}ndez-Trincado},
  {Feuillet}, {Finoguenov}, {Fofie}, {Fraser-McKelvie}, {Frinchaboy},
  {Fromenteau}, {Fu}, {Galbany}, {Garcia}, {Garc{\'\i}a-Hern{\'a}ndez}, {Garma
  Oehmichen}, {Ge}, {Geimba Maia}, {Geisler}, {Gelfand }, {Goddy},
  {Gonzalez-Perez}, {Grabowski}, {Green}, {Grier}, {Guo}, {Guy}, {Harding},
  {Hasselquist}, {Hawken}, {Hayes}, {Hearty}, {Hekker}, {Hogg}, {Holtzman},
  {Horta}, {Hou}, {Hsieh}, {Huber}, {Hunt}, {Ider Chitham}, {Imig}, {Jaber},
  {Jimenez Angel}, {Johnson}, {Jones}, {J{\"o}nsson}, {Jullo}, {Kim},
  {Kinemuchi}, {Kirkpatrick}, {Kite}, {Klaene}, {Kneib}, {Kollmeier}, {Kong},
  {Kounkel}, {Krishnarao}, {Lacerna}, {Lan}, {Lane}, {Law}, {Le Goff}, {Leung},
  {Lewis}, {Li}, {Lian}, {Lin}, {Long}, {Longa-Pe{\~n}a}, {Lundgren}, {Lyke},
  {Ted Mackereth}, {MacLeod}, {Majewski}, {Manchado}, {Maraston}, {Martini},
  {Masseron}, {Masters}, {Mathur}, {McDermid}, {Merloni}, {Merrifield},
  {M{\'e}sz{\'a}ros}, {Miglio}, {Minniti}, {Minsley}, {Miyaji}, {Mohammad},
  {Mosser}, {Mueller}, {Muna}, {Mu{\~n}oz-Guti{\'e}rrez}, {Myers}, {Nadathur},
  {Nair}, {Nandra}, {do Nascimento}, {Nevin}, {Newman}, {Nidever}, {Nitschelm},
  {Noterdaeme}, {O'Connell}, {Olmstead}, {Oravetz}, {Oravetz}, {Osorio},
  {Pace}, {Padilla}, {Palanque-Delabrouille}, {Palicio}, {Pan}, {Pan},
  {Parker}, {Paviot}, {Peirani}, {Pe{\~n}a Ram{\'r}ez}, {Penny}, {Percival},
  {Perez-Fournon}, {P{\'e}rez-R{\`a}fols}, {Petitjean}, {Pieri},
  {Pinsonneault}, {Poovelil}, {Povick}, {Prakash}, {Price-Whelan}, {Raddick},
  {Raichoor}, {Ray}, {Rembold}, {Rezaie}, {Riffel}, {Riffel}, {Rix}, {Robin},
  {Roman-Lopes}, {Rom{\'a}n-Z{\'u}{\~n}iga}, {Rose}, {Ross}, {Rossi}, {Rowland
  s}, {Rubin}, {Salvato}, {S{\'a}nchez}, {S{\'a}nchez-Menguiano},
  {S{\'a}nchez-Gallego}, {Sayres}, {Schaefer}, {Schiavon}, {Schimoia},
  {Schlafly}, {Schlegel}, {Schneider}, {Schultheis}, {Schwope}, {Seo},
  {Serenelli}, {Shafieloo}, {Shamsi}, {Shao}, {Shen}, {Shetrone}, {Shirley},
  {Silva Aguirre}, {Simon}, {Skrutskie}, {Slosar}, {Smethurst}, {Sobeck},
  {Sodi}, {Souto}, {Stark}, {Stassun}, {Steinmetz}, {Stello}, {Stermer},
  {Storchi-Bergmann}, {Streblyanska}, {Stringfellow}, {Stutz}, {Su{\'a}rez},
  {Sun}, {Taghizadeh-Popp}, {Talbot}, {Tayar}, {Thakar}, {Theriault}, {Thomas},
  {Thomas}, {Tinker}, {Tojeiro}, {Toledo}, {Tremonti}, {Troup}, {Tuttle},
  {Unda-Sanzana}, {Valentini}, {Vargas-Gonz{\'a}lez}, {Vargas-Maga{\~n}a},
  {V{\'a}zquez-Mata}, {Vivek}, {Wake}, {Wang}, {Weaver}, {Weijmans}, {Wild},
  {Wilson}, {Wilson}, {Wolthuis}, {Wood-Vasey}, {Yan}, {Yang}, {Y{\`e}che},
  {Zamora}, {Zarrouk}, {Zasowski}, {Zhang}, {Zhao}, {Zhao}, {Zheng}, {Zheng},
  {Zhu}, \& {Zou}}]{DR162020}
{Ahumada}, R., {Allende Prieto}, C., {Almeida}, A., {et~al.} 2020, \apjs, 249,
  3, \dodoi{10.3847/1538-4365/ab929e}

\bibitem[{{Angus} {et~al.}(2015){Angus}, {McQuillan}, {Foreman-Mackey},
  {Chaplin}, \& {Mazeh}}]{Angus2015}
{Angus}, R., {McQuillan}, A., {Foreman-Mackey}, D., {Chaplin}, W.~J., \&
  {Mazeh}, T. 2015, in American Astronomical Society Meeting Abstracts, Vol.
  225, American Astronomical Society Meeting Abstracts \#225, 112.04

\bibitem[{{Angus} {et~al.}(2019{\natexlab{a}}){Angus}, {Morton}, \&
  {Foreman-Mackey}}]{Angus2019b}
{Angus}, R., {Morton}, T., \& {Foreman-Mackey}, D. 2019{\natexlab{a}}, The
  Journal of Open Source Software, 4, 1469, \dodoi{10.21105/joss.01469}

\bibitem[{{Angus} {et~al.}(2019{\natexlab{b}}){Angus}, {Morton},
  {Foreman-Mackey}, {van Saders}, {Curtis}, {Kane}, {Bedell}, {Kiman}, {Hogg},
  \& {Brewer}}]{Angus2019a}
{Angus}, R., {Morton}, T.~D., {Foreman-Mackey}, D., {et~al.}
  2019{\natexlab{b}}, \aj, 158, 173, \dodoi{10.3847/1538-3881/ab3c53}

\bibitem[{{Angus} {et~al.}(2020){Angus}, {Beane}, {Price-Whelan}, {Newton},
  {Curtis}, {Berger}, {van Saders}, {Kiman}, {Foreman-Mackey}, {Lu},
  {Anderson}, \& {Faherty}}]{Angus2020_kinage}
{Angus}, R., {Beane}, A., {Price-Whelan}, A.~M., {et~al.} 2020, \aj, 160, 90,
  \dodoi{10.3847/1538-3881/ab91b2}

\bibitem[{{Astropy Collaboration} {et~al.}(2013){Astropy Collaboration},
  {Robitaille}, {Tollerud}, {Greenfield}, {Droettboom}, {Bray}, {Aldcroft},
  {Davis}, {Ginsburg}, {Price-Whelan}, {Kerzendorf}, {Conley}, {Crighton},
  {Barbary}, {Muna}, {Ferguson}, {Grollier}, {Parikh}, {Nair}, {Unther},
  {Deil}, {Woillez}, {Conseil}, {Kramer}, {Turner}, {Singer}, {Fox}, {Weaver},
  {Zabalza}, {Edwards}, {Azalee Bostroem}, {Burke}, {Casey}, {Crawford},
  {Dencheva}, {Ely}, {Jenness}, {Labrie}, {Lim}, {Pierfederici}, {Pontzen},
  {Ptak}, {Refsdal}, {Servillat}, \& {Streicher}}]{astropy:2013}
{Astropy Collaboration}, {Robitaille}, T.~P., {Tollerud}, E.~J., {et~al.} 2013,
  \aap, 558, A33, \dodoi{10.1051/0004-6361/201322068}

\bibitem[{{Aumer} \& {Binney}(2009)}]{Aumer2009}
{Aumer}, M., \& {Binney}, J.~J. 2009, \mnras, 397, 1286,
  \dodoi{10.1111/j.1365-2966.2009.15053.x}

\bibitem[{{Barbanis} \& {Woltjer}(1967)}]{Barbanis1967}
{Barbanis}, B., \& {Woltjer}, L. 1967, \apj, 150, 461, \dodoi{10.1086/149349}

\bibitem[{{Barnes}(2003)}]{Barnes2003}
{Barnes}, S.~A. 2003, \apj, 586, 464, \dodoi{10.1086/367639}

\bibitem[{{Barnes}(2007)}]{Barnes2007}
---. 2007, \apj, 669, 1167, \dodoi{10.1086/519295}

\bibitem[{{Barnes}(2010)}]{Barnes2010}
---. 2010, \apj, 722, 222, \dodoi{10.1088/0004-637X/722/1/222}

\bibitem[{{Bedell} {et~al.}(2018){Bedell}, {Bean}, {Mel{\'e}ndez}, {Spina},
  {Ram{\'\i}rez}, {Asplund}, {Alves-Brito}, {dos Santos}, {Dreizler}, {Yong},
  {Monroe}, \& {Casagrande}}]{Bedell2018}
{Bedell}, M., {Bean}, J.~L., {Mel{\'e}ndez}, J., {et~al.} 2018, \apj, 865, 68,
  \dodoi{10.3847/1538-4357/aad908}

\bibitem[{{Berger} {et~al.}(2020){Berger}, {Huber}, {Gaidos}, {van Saders}, \&
  {Weiss}}]{Berger2020}
{Berger}, T.~A., {Huber}, D., {Gaidos}, E., {van Saders}, J.~L., \& {Weiss},
  L.~M. 2020, arXiv e-prints, arXiv:2005.14671.
\newblock \doarXiv{2005.14671}

\bibitem[{Bergeron {et~al.}(1995)Bergeron, Wesemael, \&
  Beauchamp}]{Bergeron_1995}
Bergeron, P., Wesemael, F., \& Beauchamp, A. 1995, Publications of the
  Astronomical Society of the Pacific, 107, 1047, \dodoi{10.1086/133661}

\bibitem[{{Binks} {et~al.}(2020){Binks}, {Jeffries}, \& {Wright}}]{Binks2020}
{Binks}, A.~S., {Jeffries}, R.~D., \& {Wright}, N.~J. 2020, \mnras, 494, 2429,
  \dodoi{10.1093/mnras/staa909}

\bibitem[{{Bird} {et~al.}(2013){Bird}, {Kazantzidis}, {Weinberg}, {Guedes},
  {Callegari}, {Mayer}, \& {Madau}}]{Bird2013}
{Bird}, J.~C., {Kazantzidis}, S., {Weinberg}, D.~H., {et~al.} 2013, \apj, 773,
  43, \dodoi{10.1088/0004-637X/773/1/43}

\bibitem[{{Borucki} {et~al.}(2010){Borucki}, {Koch}, {Basri}, {Batalha},
  {Brown}, {Caldwell}, {Caldwell}, {Christensen-Dalsgaard}, {Cochran},
  {DeVore}, {Dunham}, {Dupree}, {Gautier}, {Geary}, {Gilliland}, {Gould},
  {Howell}, {Jenkins}, {Kondo}, {Latham}, {Marcy}, {Meibom}, {Kjeldsen},
  {Lissauer}, {Monet}, {Morrison}, {Sasselov}, {Tarter}, {Boss}, {Brownlee},
  {Owen}, {Buzasi}, {Charbonneau}, {Doyle}, {Fortney}, {Ford}, {Holman},
  {Seager}, {Steffen}, {Welsh}, {Rowe}, {Anderson}, {Buchhave}, {Ciardi},
  {Walkowicz}, {Sherry}, {Horch}, {Isaacson}, {Everett}, {Fischer}, {Torres},
  {Johnson}, {Endl}, {MacQueen}, {Bryson}, {Dotson}, {Haas}, {Kolodziejczak},
  {Van Cleve}, {Chandrasekaran}, {Twicken}, {Quintana}, {Clarke}, {Allen},
  {Li}, {Wu}, {Tenenbaum}, {Verner}, {Bruhweiler}, {Barnes}, \&
  {Prsa}}]{Borucki2010}
{Borucki}, W.~J., {Koch}, D., {Basri}, G., {et~al.} 2010, Science, 327, 977,
  \dodoi{10.1126/science.1185402}

\bibitem[{{Bovy} {et~al.}(2011){Bovy}, {Hogg}, \& {Roweis}}]{Bovy2011}
{Bovy}, J., {Hogg}, D.~W., \& {Roweis}, S.~T. 2011, Annals of Applied
  Statistics, 5, 1657, \dodoi{10.1214/10-AOAS439}

\bibitem[{{Buder} {et~al.}(2019){Buder}, {Lind}, {Ness}, {Asplund}, {Duong},
  {Lin}, {Kos}, {Casagrande}, {Casey}, {Bland-Hawthorn}, {de Silva}, {D'Orazi},
  {Freeman}, {Martell}, {Schlesinger}, {Sharma}, {Simpson}, {Zucker},
  {Zwitter}, {{\v{C}}otar}, {Dotter}, {Hayden}, {Hyde}, {Kafle}, {Lewis},
  {Nataf}, {Nordlander}, {Reid}, {Rix}, {Sk{\'u}lad{\'o}ttir}, {Stello},
  {Ting}, {Traven}, {Wyse}, \& {Galah Collaboration}}]{Buder2019}
{Buder}, S., {Lind}, K., {Ness}, M.~K., {et~al.} 2019, \aap, 624, A19,
  \dodoi{10.1051/0004-6361/201833218}

\bibitem[{{Choi} {et~al.}(2016){Choi}, {Dotter}, {Conroy}, {Cantiello},
  {Paxton}, \& {Johnson}}]{Choi2016}
{Choi}, J., {Dotter}, A., {Conroy}, C., {et~al.} 2016, \apj, 823, 102,
  \dodoi{10.3847/0004-637X/823/2/102}

\bibitem[{{Claytor} {et~al.}(2020){Claytor}, {van Saders}, {Santos},
  {Garc{\'\i}a}, {Mathur}, {Tayar}, {Pinsonneault}, \&
  {Shetrone}}]{Claytor2020}
{Claytor}, Z.~R., {van Saders}, J.~L., {Santos}, {\^A}. R.~G., {et~al.} 2020,
  \apj, 888, 43, \dodoi{10.3847/1538-4357/ab5c24}

\bibitem[{{Cui} {et~al.}(2012{\natexlab{a}}){Cui}, {Zhao}, {Chu}, {Li}, {Li},
  {Zhang}, {Su}, {Yao}, {Wang}, {Xing}, {Li}, {Zhu}, {Wang}, {Gu}, {Luo}, {Xu},
  {Zhang}, {Liu}, {Zhang}, {Yang}, {Cao}, {Chen}, {Chen}, {Chen}, {Chen},
  {Chu}, {Feng}, {Gong}, {Hou}, {Hu}, {Hu}, {Hu}, {Jia}, {Jiang}, {Jiang},
  {Jiang}, {Jin}, {Li}, {Li}, {Li}, {Liu}, {Liu}, {Lu}, {Mao}, {Men}, {Qi},
  {Qi}, {Shi}, {Tang}, {Tao}, {Wang}, {Wang}, {Wang}, {Wang}, {Wang}, {Wang},
  {Wang}, {Wang}, {Wang}, {Wang}, {Wang}, {Wang}, {Xu}, {Xu}, {Yang}, {Yu},
  {Yuan}, {Yuan}, {Zhai}, {Zhang}, {Zhang}, {Zhang}, {Zhao}, {Zhou}, {Zhou},
  {Zhu}, \& {Zou}}]{LAMOST}
{Cui}, X.-Q., {Zhao}, Y.-H., {Chu}, Y.-Q., {et~al.} 2012{\natexlab{a}},
  Research in Astronomy and Astrophysics, 12, 1197,
  \dodoi{10.1088/1674-4527/12/9/003}

\bibitem[{{Cui} {et~al.}(2012{\natexlab{b}}){Cui}, {Zhao}, {Chu}, {Li}, {Li},
  {Zhang}, {Su}, {Yao}, {Wang}, {Xing}, {Li}, {Zhu}, {Wang}, {Gu}, {Luo}, {Xu},
  {Zhang}, {Liu}, {Zhang}, {Yang}, {Cao}, {Chen}, {Chen}, {Chen}, {Chen},
  {Chu}, {Feng}, {Gong}, {Hou}, {Hu}, {Hu}, {Hu}, {Jia}, {Jiang}, {Jiang},
  {Jiang}, {Jin}, {Li}, {Li}, {Li}, {Liu}, {Liu}, {Lu}, {Mao}, {Men}, {Qi},
  {Qi}, {Shi}, {Tang}, {Tao}, {Wang}, {Wang}, {Wang}, {Wang}, {Wang}, {Wang},
  {Wang}, {Wang}, {Wang}, {Wang}, {Wang}, {Wang}, {Xu}, {Xu}, {Yang}, {Yu},
  {Yuan}, {Yuan}, {Zhai}, {Zhang}, {Zhang}, {Zhang}, {Zhao}, {Zhou}, {Zhou},
  {Zhu}, \& {Zou}}]{Cui2012}
---. 2012{\natexlab{b}}, Research in Astronomy and Astrophysics, 12, 1197,
  \dodoi{10.1088/1674-4527/12/9/003}

\bibitem[{{Cummings} {et~al.}(2018){Cummings}, {Kalirai}, {Tremblay},
  {Ramirez-Ruiz}, \& {Choi}}]{Cummings2018}
{Cummings}, J.~D., {Kalirai}, J.~S., {Tremblay}, P.~E., {Ramirez-Ruiz}, E., \&
  {Choi}, J. 2018, \apj, 866, 21, \dodoi{10.3847/1538-4357/aadfd6}

\bibitem[{{Curtis} {et~al.}(2019){Curtis}, {Ag{\"u}eros}, {Douglas}, \&
  {Meibom}}]{Curtis2019}
{Curtis}, J.~L., {Ag{\"u}eros}, M.~A., {Douglas}, S.~T., \& {Meibom}, S. 2019,
  \apj, 879, 49, \dodoi{10.3847/1538-4357/ab2393}

\bibitem[{{Curtis} {et~al.}(2020){Curtis}, {Ag{\"u}eros}, {Matt}, {Covey},
  {Douglas}, {Angus}, {Saar}, {Cody}, {Vanderburg}, {Law}, {Kraus}, {Latham},
  {Baranec}, {Riddle}, {Ziegler}, {Lund}, {Torres}, {Meibom}, {Aguirre}, \&
  {Wright}}]{Curtis2020}
{Curtis}, J.~L., {Ag{\"u}eros}, M.~A., {Matt}, S.~P., {et~al.} 2020, \apj, 904,
  140, \dodoi{10.3847/1538-4357/abbf58}

\bibitem[{{David} {et~al.}(2020){David}, {Contardo}, {Sandoval}, {Angus},
  {Yuxi}, {Lu}, {Bedell}, {Curtis}, {Foreman-Mackey}, {Fulton}, {Grunblatt}, \&
  {Petigura}}]{David2020}
{David}, T.~J., {Contardo}, G., {Sandoval}, A., {et~al.} 2020, arXiv e-prints,
  arXiv:2011.09894.
\newblock \doarXiv{2011.09894}

\bibitem[{{Dotter}(2016)}]{Dotter2016}
{Dotter}, A. 2016, \apjs, 222, 8, \dodoi{10.3847/0067-0049/222/1/8}

\bibitem[{{Douglas} {et~al.}(2016){Douglas}, {Ag{\"u}eros}, {Covey}, {Cargile},
  {Barclay}, {Cody}, {Howell}, \& {Kopytova}}]{Douglas2016}
{Douglas}, S.~T., {Ag{\"u}eros}, M.~A., {Covey}, K.~R., {et~al.} 2016, \apj,
  822, 47, \dodoi{10.3847/0004-637X/822/1/47}

\bibitem[{{Douglas} {et~al.}(2017){Douglas}, {Ag{\"u}eros}, {Covey}, \&
  {Kraus}}]{Douglas2017}
{Douglas}, S.~T., {Ag{\"u}eros}, M.~A., {Covey}, K.~R., \& {Kraus}, A. 2017,
  \apj, 842, 83, \dodoi{10.3847/1538-4357/aa6e52}

\bibitem[{{Douglas} {et~al.}(2019){Douglas}, {Curtis}, {Ag{\"u}eros},
  {Cargile}, {Brewer}, {Meibom}, \& {Jansen}}]{Douglas2019}
{Douglas}, S.~T., {Curtis}, J.~L., {Ag{\"u}eros}, M.~A., {et~al.} 2019, \apj,
  879, 100, \dodoi{10.3847/1538-4357/ab2468}

\bibitem[{{Gaia Collaboration} {et~al.}(2016){Gaia Collaboration}, {Prusti},
  {de Bruijne}, {Brown}, {Vallenari}, {Babusiaux}, {Bailer-Jones}, {Bastian},
  {Biermann}, {Evans}, {Eyer}, {Jansen}, {Jordi}, {Klioner}, {Lammers},
  {Lindegren}, {Luri}, {Mignard}, {Milligan}, {Panem}, {Poinsignon},
  {Pourbaix}, {Randich}, {Sarri}, {Sartoretti}, {Siddiqui}, {Soubiran},
  {Valette}, {van Leeuwen}, {Walton}, {Aerts}, {Arenou}, {Cropper}, {Drimmel},
  {H{\o}g}, {Katz}, {Lattanzi}, {O'Mullane}, {Grebel}, {Holland}, {Huc},
  {Passot}, {Bramante}, {Cacciari}, {Casta{\~n}eda}, {Chaoul}, {Cheek}, {De
  Angeli}, {Fabricius}, {Guerra}, {Hern{\'a}ndez}, {Jean-Antoine-Piccolo},
  {Masana}, {Messineo}, {Mowlavi}, {Nienartowicz}, {Ord{\'o}{\~n}ez-Blanco},
  {Panuzzo}, {Portell}, {Richards}, {Riello}, {Seabroke}, {Tanga},
  {Th{\'e}venin}, {Torra}, {Els}, {Gracia-Abril}, {Comoretto},
  {Garcia-Reinaldos}, {Lock}, {Mercier}, {Altmann}, {Andrae}, {Astraatmadja},
  {Bellas-Velidis}, {Benson}, {Berthier}, {Blomme}, {Busso}, {Carry},
  {Cellino}, {Clementini}, {Cowell}, {Creevey}, {Cuypers}, {Davidson}, {De
  Ridder}, {de Torres}, {Delchambre}, {Dell'Oro}, {Ducourant}, {Fr{\'e}mat},
  {Garc{\'\i}a-Torres}, {Gosset}, {Halbwachs}, {Hambly}, {Harrison}, {Hauser},
  {Hestroffer}, {Hodgkin}, {Huckle}, {Hutton}, {Jasniewicz}, {Jordan},
  {Kontizas}, {Korn}, {Lanzafame}, {Manteiga}, {Moitinho}, {Muinonen},
  {Osinde}, {Pancino}, {Pauwels}, {Petit}, {Recio-Blanco}, {Robin}, {Sarro},
  {Siopis}, {Smith}, {Smith}, {Sozzetti}, {Thuillot}, {van Reeven}, {Viala},
  {Abbas}, {Abreu Aramburu}, {Accart}, {Aguado}, {Allan}, {Allasia},
  {Altavilla}, {{\'A}lvarez}, {Alves}, {Anderson}, {Andrei}, {Anglada Varela},
  {Antiche}, {Antoja}, {Ant{\'o}n}, {Arcay}, {Atzei}, {Ayache}, {Bach},
  {Baker}, {Balaguer-N{\'u}{\~n}ez}, {Barache}, {Barata}, {Barbier}, {Barblan},
  {Baroni}, {Barrado y Navascu{\'e}s}, {Barros}, {Barstow}, {Becciani},
  {Bellazzini}, {Bellei}, {Bello Garc{\'\i}a}, {Belokurov}, {Bendjoya},
  {Berihuete}, {Bianchi}, {Bienaym{\'e}}, {Billebaud}, {Blagorodnova},
  {Blanco-Cuaresma}, {Boch}, {Bombrun}, {Borrachero}, {Bouquillon}, {Bourda},
  {Bouy}, {Bragaglia}, {Breddels}, {Brouillet}, {Br{\"u}semeister},
  {Bucciarelli}, {Budnik}, {Burgess}, {Burgon}, {Burlacu}, {Busonero}, {Buzzi},
  {Caffau}, {Cambras}, {Campbell}, {Cancelliere}, {Cantat-Gaudin}, {Carlucci},
  {Carrasco}, {Castellani}, {Charlot}, {Charnas}, {Charvet}, {Chassat},
  {Chiavassa}, {Clotet}, {Cocozza}, {Collins}, {Collins}, {Costigan}, {Crifo},
  {Cross}, {Crosta}, {Crowley}, {Dafonte}, {Damerdji}, {Dapergolas}, {David},
  {David}, {De Cat}, {de Felice}, {de Laverny}, {De Luise}, {De March}, {de
  Martino}, {de Souza}, {Debosscher}, {del Pozo}, {Delbo}, {Delgado},
  {Delgado}, {di Marco}, {Di Matteo}, {Diakite}, {Distefano}, {Dolding}, {Dos
  Anjos}, {Drazinos}, {Dur{\'a}n}, {Dzigan}, {Ecale}, {Edvardsson}, {Enke},
  {Erdmann}, {Escolar}, {Espina}, {Evans}, {Eynard Bontemps}, {Fabre},
  {Fabrizio}, {Faigler}, {Falc{\~a}o}, {Farr{\`a}s Casas}, {Faye}, {Federici},
  {Fedorets}, {Fern{\'a}ndez-Hern{\'a}ndez}, {Fernique}, {Fienga}, {Figueras},
  {Filippi}, {Findeisen}, {Fonti}, {Fouesneau}, {Fraile}, {Fraser}, {Fuchs},
  {Furnell}, {Gai}, {Galleti}, {Galluccio}, {Garabato}, {Garc{\'\i}a-Sedano},
  {Gar{\'e}}, {Garofalo}, {Garralda}, {Gavras}, {Gerssen}, {Geyer}, {Gilmore},
  {Girona}, {Giuffrida}, {Gomes}, {Gonz{\'a}lez-Marcos},
  {Gonz{\'a}lez-N{\'u}{\~n}ez}, {Gonz{\'a}lez-Vidal}, {Granvik}, {Guerrier},
  {Guillout}, {Guiraud}, {G{\'u}rpide}, {Guti{\'e}rrez-S{\'a}nchez}, {Guy},
  {Haigron}, {Hatzidimitriou}, {Haywood}, {Heiter}, {Helmi}, {Hobbs},
  {Hofmann}, {Holl}, {Holland }, {Hunt}, {Hypki}, {Icardi}, {Irwin}, {Jevardat
  de Fombelle}, {Jofr{\'e}}, {Jonker}, {Jorissen}, {Julbe}, {Karampelas},
  {Kochoska}, {Kohley}, {Kolenberg}, {Kontizas}, {Koposov}, {Kordopatis},
  {Koubsky}, {Kowalczyk}, {Krone-Martins}, {Kudryashova}, {Kull}, {Bachchan},
  {Lacoste-Seris}, {Lanza}, {Lavigne}, {Le Poncin-Lafitte}, {Lebreton},
  {Lebzelter}, {Leccia}, {Leclerc}, {Lecoeur-Taibi}, {Lemaitre}, {Lenhardt},
  {Leroux}, {Liao}, {Licata}, {Lindstr{\o}m}, {Lister}, {Livanou}, {Lobel},
  {L{\"o}ffler}, {L{\'o}pez}, {Lopez-Lozano}, {Lorenz}, {Loureiro},
  {MacDonald}, {Magalh{\~a}es Fernandes}, {Managau}, {Mann}, {Mantelet},
  {Marchal}, {Marchant}, {Marconi}, {Marie}, {Marinoni}, {Marrese},
  {Marschalk{\'o}}, {Marshall}, {Mart{\'\i}n-Fleitas}, {Martino}, {Mary},
  {Matijevi{\v{c}}}, {Mazeh}, {McMillan}, {Messina}, {Mestre}, {Michalik},
  {Millar}, {Miranda}, {Molina}, {Molinaro}, {Molinaro}, {Moln{\'a}r},
  {Moniez}, {Montegriffo}, {Monteiro}, {Mor}, {Mora}, {Morbidelli}, {Morel},
  {Morgenthaler}, {Morley}, {Morris}, {Mulone}, {Muraveva}, {Musella},
  {Narbonne}, {Nelemans}, {Nicastro}, {Noval}, {Ord{\'e}novic},
  {Ordieres-Mer{\'e}}, {Osborne}, {Pagani}, {Pagano}, {Pailler}, {Palacin},
  {Palaversa}, {Parsons}, {Paulsen}, {Pecoraro}, {Pedrosa}, {Pentik{\"a}inen},
  {Pereira}, {Pichon}, {Piersimoni}, {Pineau}, {Plachy}, {Plum}, {Poujoulet},
  {Pr{\v{s}}a}, {Pulone}, {Ragaini}, {Rago}, {Rambaux}, {Ramos-Lerate},
  {Ranalli}, {Rauw}, {Read}, {Regibo}, {Renk}, {Reyl{\'e}}, {Ribeiro},
  {Rimoldini}, {Ripepi}, {Riva}, {Rixon}, {Roelens}, {Romero-G{\'o}mez},
  {Rowell}, {Royer}, {Rudolph}, {Ruiz-Dern}, {Sadowski}, {Sagrist{\`a}
  Sell{\'e}s}, {Sahlmann}, {Salgado}, {Salguero}, {Sarasso}, {Savietto},
  {Schnorhk}, {Schultheis}, {Sciacca}, {Segol}, {Segovia}, {Segransan},
  {Serpell}, {Shih}, {Smareglia}, {Smart}, {Smith}, {Solano}, {Solitro},
  {Sordo}, {Soria Nieto}, {Souchay}, {Spagna}, {Spoto}, {Stampa}, {Steele},
  {Steidelm{\"u}ller}, {Stephenson}, {Stoev}, {Suess}, {S{\"u}veges}, {Surdej},
  {Szabados}, {Szegedi-Elek}, {Tapiador}, {Taris}, {Tauran}, {Taylor},
  {Teixeira}, {Terrett}, {Tingley}, {Trager}, {Turon}, {Ulla}, {Utrilla},
  {Valentini}, {van Elteren}, {Van Hemelryck}, {van Leeuwen}, {Varadi},
  {Vecchiato}, {Veljanoski}, {Via}, {Vicente}, {Vogt}, {Voss}, {Votruba},
  {Voutsinas}, {Walmsley}, {Weiler}, {Weingrill}, {Werner}, {Wevers},
  {Whitehead}, {Wyrzykowski}, {Yoldas}, {{\v{Z}}erjal}, {Zucker}, {Zurbach},
  {Zwitter}, {Alecu}, {Allen}, {Allende Prieto}, {Amorim},
  {Anglada-Escud{\'e}}, {Arsenijevic}, {Azaz}, {Balm}, {Beck}, {Bernstein},
  {Bigot}, {Bijaoui}, {Blasco}, {Bonfigli}, {Bono}, {Boudreault}, {Bressan},
  {Brown}, {Brunet}, {Bunclark}, {Buonanno}, {Butkevich}, {Carret}, {Carrion},
  {Chemin}, {Ch{\'e}reau}, {Corcione}, {Darmigny}, {de Boer}, {de Teodoro}, {de
  Zeeuw}, {Delle Luche}, {Domingues}, {Dubath}, {Fodor}, {Fr{\'e}zouls},
  {Fries}, {Fustes}, {Fyfe}, {Gallardo}, {Gallegos}, {Gardiol}, {Gebran},
  {Gomboc}, {G{\'o}mez}, {Grux}, {Gueguen}, {Heyrovsky}, {Hoar}, {Iannicola},
  {Isasi Parache}, {Janotto}, {Joliet}, {Jonckheere}, {Keil}, {Kim},
  {Klagyivik}, {Klar}, {Knude}, {Kochukhov}, {Kolka}, {Kos}, {Kutka}, {Lainey},
  {LeBouquin}, {Liu}, {Loreggia}, {Makarov}, {Marseille}, {Martayan},
  {Martinez-Rubi}, {Massart}, {Meynadier}, {Mignot}, {Munari}, {Nguyen},
  {Nordlander}, {Ocvirk}, {O'Flaherty}, {Olias Sanz}, {Ortiz}, {Osorio},
  {Oszkiewicz}, {Ouzounis}, {Palmer}, {Park}, {Pasquato}, {Peltzer}, {Peralta},
  {P{\'e}turaud}, {Pieniluoma}, {Pigozzi}, {Poels}, {Prat}, {Prod'homme},
  {Raison}, {Rebordao}, {Risquez}, {Rocca-Volmerange}, {Rosen}, {Ruiz-Fuertes},
  {Russo}, {Sembay}, {Serraller Vizcaino}, {Short}, {Siebert}, {Silva},
  {Sinachopoulos}, {Slezak}, {Soffel}, {Sosnowska}, {Strai{\v{z}}ys}, {ter
  Linden}, {Terrell}, {Theil}, {Tiede}, {Troisi}, {Tsalmantza}, {Tur},
  {Vaccari}, {Vachier}, {Valles}, {Van Hamme}, {Veltz}, {Virtanen}, {Wallut},
  {Wichmann}, {Wilkinson}, {Ziaeepour}, \& {Zschocke}}]{Prusti2016}
{Gaia Collaboration}, {Prusti}, T., {de Bruijne}, J.~H.~J., {et~al.} 2016,
  \aap, 595, A1, \dodoi{10.1051/0004-6361/201629272}

\bibitem[{{Gaia Collaboration} {et~al.}(2018){Gaia Collaboration}, {Brown},
  {Vallenari}, {Prusti}, {de Bruijne}, {Babusiaux}, {Bailer-Jones}, {Biermann},
  {Evans}, {Eyer}, {Jansen}, {Jordi}, {Klioner}, {Lammers}, {Lindegren},
  {Luri}, {Mignard}, {Panem}, {Pourbaix}, {Randich}, {Sartoretti}, {Siddiqui},
  {Soubiran}, {van Leeuwen}, {Walton}, {Arenou}, {Bastian}, {Cropper},
  {Drimmel}, {Katz}, {Lattanzi}, {Bakker}, {Cacciari}, {Casta{\~n}eda},
  {Chaoul}, {Cheek}, {De Angeli}, {Fabricius}, {Guerra}, {Holl}, {Masana},
  {Messineo}, {Mowlavi}, {Nienartowicz}, {Panuzzo}, {Portell}, {Riello},
  {Seabroke}, {Tanga}, {Th{\'e}venin}, {Gracia-Abril}, {Comoretto},
  {Garcia-Reinaldos}, {Teyssier}, {Altmann}, {Andrae}, {Audard},
  {Bellas-Velidis}, {Benson}, {Berthier}, {Blomme}, {Burgess}, {Busso},
  {Carry}, {Cellino}, {Clementini}, {Clotet}, {Creevey}, {Davidson}, {De
  Ridder}, {Delchambre}, {Dell'Oro}, {Ducourant},
  {Fern{\'a}ndez-Hern{\'a}ndez}, {Fouesneau}, {Fr{\'e}mat}, {Galluccio},
  {Garc{\'\i}a-Torres}, {Gonz{\'a}lez-N{\'u}{\~n}ez}, {Gonz{\'a}lez-Vidal},
  {Gosset}, {Guy}, {Halbwachs}, {Hambly}, {Harrison}, {Hern{\'a}ndez},
  {Hestroffer}, {Hodgkin}, {Hutton}, {Jasniewicz}, {Jean-Antoine-Piccolo},
  {Jordan}, {Korn}, {Krone-Martins}, {Lanzafame}, {Lebzelter}, {L{\"o}ffler},
  {Manteiga}, {Marrese}, {Mart{\'\i}n-Fleitas}, {Moitinho}, {Mora}, {Muinonen},
  {Osinde}, {Pancino}, {Pauwels}, {Petit}, {Recio-Blanco}, {Richards},
  {Rimoldini}, {Robin}, {Sarro}, {Siopis}, {Smith}, {Sozzetti}, {S{\"u}veges},
  {Torra}, {van Reeven}, {Abbas}, {Abreu Aramburu}, {Accart}, {Aerts},
  {Altavilla}, {{\'A}lvarez}, {Alvarez}, {Alves}, {Anderson}, {Andrei},
  {Anglada Varela}, {Antiche}, {Antoja}, {Arcay}, {Astraatmadja}, {Bach},
  {Baker}, {Balaguer-N{\'u}{\~n}ez}, {Balm}, {Barache}, {Barata}, {Barbato},
  {Barblan}, {Barklem}, {Barrado}, {Barros}, {Barstow}, {Bartholom{\'e}
  Mu{\~n}oz}, {Bassilana}, {Becciani}, {Bellazzini}, {Berihuete}, {Bertone},
  {Bianchi}, {Bienaym{\'e}}, {Blanco-Cuaresma}, {Boch}, {Boeche}, {Bombrun},
  {Borrachero}, {Bossini}, {Bouquillon}, {Bourda}, {Bragaglia}, {Bramante},
  {Breddels}, {Bressan}, {Brouillet}, {Br{\"u}semeister}, {Brugaletta},
  {Bucciarelli}, {Burlacu}, {Busonero}, {Butkevich}, {Buzzi}, {Caffau},
  {Cancelliere}, {Cannizzaro}, {Cantat-Gaudin}, {Carballo}, {Carlucci},
  {Carrasco}, {Casamiquela}, {Castellani}, {Castro-Ginard}, {Charlot},
  {Chemin}, {Chiavassa}, {Cocozza}, {Costigan}, {Cowell}, {Crifo}, {Crosta},
  {Crowley}, {Cuypers}, {Dafonte}, {Damerdji}, {Dapergolas}, {David}, {David},
  {de Laverny}, {De Luise}, {De March}, {de Martino}, {de Souza}, {de Torres},
  {Debosscher}, {del Pozo}, {Delbo}, {Delgado}, {Delgado}, {Di Matteo},
  {Diakite}, {Diener}, {Distefano}, {Dolding}, {Drazinos}, {Dur{\'a}n},
  {Edvardsson}, {Enke}, {Eriksson}, {Esquej}, {Eynard Bontemps}, {Fabre},
  {Fabrizio}, {Faigler}, {Falc{\~a}o}, {Farr{\`a}s Casas}, {Federici},
  {Fedorets}, {Fernique}, {Figueras}, {Filippi}, {Findeisen}, {Fonti},
  {Fraile}, {Fraser}, {Fr{\'e}zouls}, {Gai}, {Galleti}, {Garabato},
  {Garc{\'\i}a-Sedano}, {Garofalo}, {Garralda}, {Gavel}, {Gavras}, {Gerssen},
  {Geyer}, {Giacobbe}, {Gilmore}, {Girona}, {Giuffrida}, {Glass}, {Gomes},
  {Granvik}, {Gueguen}, {Guerrier}, {Guiraud}, {Guti{\'e}rrez-S{\'a}nchez},
  {Haigron}, {Hatzidimitriou}, {Hauser}, {Haywood}, {Heiter}, {Helmi}, {Heu},
  {Hilger}, {Hobbs}, {Hofmann}, {Holland}, {Huckle}, {Hypki}, {Icardi},
  {Jan{\ss}en}, {Jevardat de Fombelle}, {Jonker}, {Juh{\'a}sz}, {Julbe},
  {Karampelas}, {Kewley}, {Klar}, {Kochoska}, {Kohley}, {Kolenberg},
  {Kontizas}, {Kontizas}, {Koposov}, {Kordopatis}, {Kostrzewa-Rutkowska},
  {Koubsky}, {Lambert}, {Lanza}, {Lasne}, {Lavigne}, {Le Fustec}, {Le
  Poncin-Lafitte}, {Lebreton}, {Leccia}, {Leclerc}, {Lecoeur-Taibi},
  {Lenhardt}, {Leroux}, {Liao}, {Licata}, {Lindstr{\o}m}, {Lister}, {Livanou},
  {Lobel}, {L{\'o}pez}, {Managau}, {Mann}, {Mantelet}, {Marchal}, {Marchant},
  {Marconi}, {Marinoni}, {Marschalk{\'o}}, {Marshall}, {Martino}, {Marton},
  {Mary}, {Massari}, {Matijevi{\v{c}}}, {Mazeh}, {McMillan}, {Messina},
  {Michalik}, {Millar}, {Molina}, {Molinaro}, {Moln{\'a}r}, {Montegriffo},
  {Mor}, {Morbidelli}, {Morel}, {Morris}, {Mulone}, {Muraveva}, {Musella},
  {Nelemans}, {Nicastro}, {Noval}, {O'Mullane}, {Ord{\'e}novic},
  {Ord{\'o}{\~n}ez-Blanco}, {Osborne}, {Pagani}, {Pagano}, {Pailler},
  {Palacin}, {Palaversa}, {Panahi}, {Pawlak}, {Piersimoni}, {Pineau}, {Plachy},
  {Plum}, {Poggio}, {Poujoulet}, {Pr{\v{s}}a}, {Pulone}, {Racero}, {Ragaini},
  {Rambaux}, {Ramos-Lerate}, {Regibo}, {Reyl{\'e}}, {Riclet}, {Ripepi}, {Riva},
  {Rivard}, {Rixon}, {Roegiers}, {Roelens}, {Romero-G{\'o}mez}, {Rowell},
  {Royer}, {Ruiz-Dern}, {Sadowski}, {Sagrist{\`a} Sell{\'e}s}, {Sahlmann},
  {Salgado}, {Salguero}, {Sanna}, {Santana-Ros}, {Sarasso}, {Savietto},
  {Schultheis}, {Sciacca}, {Segol}, {Segovia}, {S{\'e}gransan}, {Shih},
  {Siltala}, {Silva}, {Smart}, {Smith}, {Solano}, {Solitro}, {Sordo}, {Soria
  Nieto}, {Souchay}, {Spagna}, {Spoto}, {Stampa}, {Steele},
  {Steidelm{\"u}ller}, {Stephenson}, {Stoev}, {Suess}, {Surdej}, {Szabados},
  {Szegedi-Elek}, {Tapiador}, {Taris}, {Tauran}, {Taylor}, {Teixeira},
  {Terrett}, {Teyssand ier}, {Thuillot}, {Titarenko}, {Torra Clotet}, {Turon},
  {Ulla}, {Utrilla}, {Uzzi}, {Vaillant}, {Valentini}, {Valette}, {van Elteren},
  {Van Hemelryck}, {van Leeuwen}, {Vaschetto}, {Vecchiato}, {Veljanoski},
  {Viala}, {Vicente}, {Vogt}, {von Essen}, {Voss}, {Votruba}, {Voutsinas},
  {Walmsley}, {Weiler}, {Wertz}, {Wevers}, {Wyrzykowski}, {Yoldas},
  {{\v{Z}}erjal}, {Ziaeepour}, {Zorec}, {Zschocke}, {Zucker}, {Zurbach}, \&
  {Zwitter}}]{Brown2018}
{Gaia Collaboration}, {Brown}, A.~G.~A., {Vallenari}, A., {et~al.} 2018, \aap,
  616, A1, \dodoi{10.1051/0004-6361/201833051}

\bibitem[{{Garc{\'\i}a} {et~al.}(2014){Garc{\'\i}a}, {Ceillier}, {Salabert},
  {Mathur}, {van Saders}, {Pinsonneault}, {Ballot}, {Beck}, {Bloemen},
  {Campante}, {Davies}, {do Nascimento}, {Mathis}, {Metcalfe}, {Nielsen},
  {Su{\'a}rez}, {Chaplin}, {Jim{\'e}nez}, \& {Karoff}}]{Garcia2014}
{Garc{\'\i}a}, R.~A., {Ceillier}, T., {Salabert}, D., {et~al.} 2014, \aap, 572,
  A34, \dodoi{10.1051/0004-6361/201423888}

\bibitem[{{Gentile Fusillo} {et~al.}(2019){Gentile Fusillo}, {Tremblay},
  {G{\"a}nsicke}, {Manser}, {Cunningham}, {Cukanovaite}, {Hollands}, {Marsh},
  {Raddi}, {Jordan}, {Toonen}, {Geier}, {Barstow}, \& {Cummings}}]{gaiawd}
{Gentile Fusillo}, N.~P., {Tremblay}, P.-E., {G{\"a}nsicke}, B.~T., {et~al.}
  2019, \mnras, 482, 4570, \dodoi{10.1093/mnras/sty3016}

\bibitem[{{Green}(2018)}]{Green2018}
{Green}, G. 2018, The Journal of Open Source Software, 3, 695,
  \dodoi{10.21105/joss.00695}

\bibitem[{{Holmberg} {et~al.}(2007){Holmberg}, {Nordstr{\"o}m}, \&
  {Andersen}}]{Holmberg2007}
{Holmberg}, J., {Nordstr{\"o}m}, B., \& {Andersen}, J. 2007, \aap, 475, 519,
  \dodoi{10.1051/0004-6361:20077221}

\bibitem[{{Holmberg} {et~al.}(2009){Holmberg}, {Nordstr{\"o}m}, \&
  {Andersen}}]{Holmberg2009}
---. 2009, \aap, 501, 941, \dodoi{10.1051/0004-6361/200811191}

\bibitem[{Hunter(2007)}]{hunter2007matplotlib}
Hunter, J.~D. 2007, Computing in science \& engineering, 9, 90

\bibitem[{{Janes} \& {Phelps}(1994)}]{Janes1994}
{Janes}, K.~A., \& {Phelps}, R.~L. 1994, \aj, 108, 1773, \dodoi{10.1086/117192}

\bibitem[{{Janes} {et~al.}(1988){Janes}, {Tilley}, \& {Lynga}}]{Janes1988}
{Janes}, K.~A., {Tilley}, C., \& {Lynga}, G. 1988, \aj, 95, 771,
  \dodoi{10.1086/114676}

\bibitem[{{Kawaler}(1988)}]{Kawaler1988}
{Kawaler}, S.~D. 1988, \apj, 333, 236, \dodoi{10.1086/166740}

\bibitem[{{Kiman} {et~al.}(2020){Kiman}, {Xu}, {Faherty}, {Angus}, {Casewell},
  {Gagn{\'e}}, \& {Cruz}}]{Kiman2020}
{Kiman}, R., {Xu}, S., {Faherty}, J.~K., {et~al.} 2020, \aj

\bibitem[{Kluyver {et~al.}(2016)Kluyver, Ragan-Kelley, P{\'e}rez, Granger,
  Bussonnier, Frederic, Kelley, Hamrick, Grout, Corlay, Ivanov, Avila, Abdalla,
  \& Willing}]{Kluyver:2016aa}
Kluyver, T., Ragan-Kelley, B., P{\'e}rez, F., {et~al.} 2016, in Positioning and
  Power in Academic Publishing: Players, Agents and Agendas, ed. F.~Loizides \&
  B.~Schmidt, IOS Press, 87 -- 90

\bibitem[{{Lacey}(1984)}]{Lacey1984}
{Lacey}, C.~G. 1984, \mnras, 208, 687, \dodoi{10.1093/mnras/208.4.687}

\bibitem[{{Lu} {et~al.}(2020){Lu}, {Angus}, {Ag{\"u}eros}, {Blancato}, {Ness},
  {Rowland}, {Curtis}, \& {Grunblatt}}]{Lu2020}
{Lu}, Y.~L., {Angus}, R., {Ag{\"u}eros}, M.~A., {et~al.} 2020, \aj, 160, 168,
  \dodoi{10.3847/1538-3881/abada4}

\bibitem[{{Majewski} {et~al.}(2017){Majewski}, {Schiavon}, {Frinchaboy},
  {Allende Prieto}, {Barkhouser}, {Bizyaev}, {Blank}, {Brunner}, {Burton},
  {Carrera}, {Chojnowski}, {Cunha}, {Epstein}, {Fitzgerald}, {Garc{\'\i}a
  P{\'e}rez}, {Hearty}, {Henderson}, {Holtzman}, {Johnson}, {Lam}, {Lawler},
  {Maseman}, {M{\'e}sz{\'a}ros}, {Nelson}, {Nguyen}, {Nidever}, {Pinsonneault},
  {Shetrone}, {Smee}, {Smith}, {Stolberg}, {Skrutskie}, {Walker}, {Wilson},
  {Zasowski}, {Anders}, {Basu}, {Beland}, {Blanton}, {Bovy}, {Brownstein},
  {Carlberg}, {Chaplin}, {Chiappini}, {Eisenstein}, {Elsworth}, {Feuillet},
  {Fleming}, {Galbraith-Frew}, {Garc{\'\i}a}, {Garc{\'\i}a-Hern{\'a}ndez},
  {Gillespie}, {Girardi}, {Gunn}, {Hasselquist}, {Hayden}, {Hekker}, {Ivans},
  {Kinemuchi}, {Klaene}, {Mahadevan}, {Mathur}, {Mosser}, {Muna}, {Munn},
  {Nichol}, {O'Connell}, {Parejko}, {Robin}, {Rocha-Pinto}, {Schultheis},
  {Serenelli}, {Shane}, {Silva Aguirre}, {Sobeck}, {Thompson}, {Troup},
  {Weinberg}, \& {Zamora}}]{Majewski2017}
{Majewski}, S.~R., {Schiavon}, R.~P., {Frinchaboy}, P.~M., {et~al.} 2017, \aj,
  154, 94, \dodoi{10.3847/1538-3881/aa784d}

\bibitem[{{Mamajek} \& {Hillenbrand}(2008)}]{Mamajek2008}
{Mamajek}, E.~E., \& {Hillenbrand}, L.~A. 2008, \apj, 687, 1264,
  \dodoi{10.1086/591785}

\bibitem[{{Martig} {et~al.}(2014){Martig}, {Minchev}, \& {Flynn}}]{Martig2014}
{Martig}, M., {Minchev}, I., \& {Flynn}, C. 2014, \mnras, 443, 2452,
  \dodoi{10.1093/mnras/stu1322}

\bibitem[{{Matt} {et~al.}(2015){Matt}, {Brun}, {Baraffe}, {Bouvier}, \&
  {Chabrier}}]{matt2015}
{Matt}, S.~P., {Brun}, A.~S., {Baraffe}, I., {Bouvier}, J., \& {Chabrier}, G.
  2015, \apjl, 799, L23, \dodoi{10.1088/2041-8205/799/2/L23}

\bibitem[{{McQuillan} {et~al.}(2014){McQuillan}, {Mazeh}, \&
  {Aigrain}}]{McQuillan2014}
{McQuillan}, A., {Mazeh}, T., \& {Aigrain}, S. 2014, \apjs, 211, 24,
  \dodoi{10.1088/0067-0049/211/2/24}

\bibitem[{{Meibom} {et~al.}(2015){Meibom}, {Barnes}, {Platais}, {Gilliland},
  {Latham}, \& {Mathieu}}]{Meibom2015}
{Meibom}, S., {Barnes}, S.~A., {Platais}, I., {et~al.} 2015, \nat, 517, 589,
  \dodoi{10.1038/nature14118}

\bibitem[{{Meibom} {et~al.}(2009){Meibom}, {Mathieu}, \&
  {Stassun}}]{Meibom2009}
{Meibom}, S., {Mathieu}, R.~D., \& {Stassun}, K.~G. 2009, \apj, 695, 679,
  \dodoi{10.1088/0004-637X/695/1/679}

\bibitem[{{Meibom} {et~al.}(2011){Meibom}, {Barnes}, {Latham}, {Batalha},
  {Borucki}, {Koch}, {Basri}, {Walkowicz}, {Janes}, {Jenkins}, {Van Cleve},
  {Haas}, {Bryson}, {Dupree}, {Furesz}, {Szentgyorgyi}, {Buchhave}, {Clarke},
  {Twicken}, \& {Quintana}}]{Meibom2011}
{Meibom}, S., {Barnes}, S.~A., {Latham}, D.~W., {et~al.} 2011, \apjl, 733, L9,
  \dodoi{10.1088/2041-8205/733/1/L9}

\bibitem[{{Ness} {et~al.}(2019){Ness}, {Johnston}, {Blancato}, {Rix}, {Beane},
  {Bird}, \& {Hawkins}}]{Ness2019}
{Ness}, M.~K., {Johnston}, K.~V., {Blancato}, K., {et~al.} 2019, \apj, 883,
  177, \dodoi{10.3847/1538-4357/ab3e3c}

\bibitem[{{Nordstr{\"o}m} {et~al.}(2004){Nordstr{\"o}m}, {Mayor}, {Andersen},
  {Holmberg}, {Pont}, {J{\o}rgensen}, {Olsen}, {Udry}, \&
  {Mowlavi}}]{Nordstrom2004}
{Nordstr{\"o}m}, B., {Mayor}, M., {Andersen}, J., {et~al.} 2004, \aap, 418,
  989, \dodoi{10.1051/0004-6361:20035959}

\bibitem[{{Ochsenbein} {et~al.}(2000){Ochsenbein}, {Bauer}, \&
  {Marcout}}]{vizier}
{Ochsenbein}, F., {Bauer}, P., \& {Marcout}, J. 2000, \aaps, 143, 23,
  \dodoi{10.1051/aas:2000169}

\bibitem[{Oliphant(2006)}]{oliphant2006guide}
Oliphant, T.~E. 2006, A guide to NumPy, Vol.~1 (Trelgol Publishing USA)

\bibitem[{pandas~development team(2020)}]{reback2020pandas}
pandas~development team, T. 2020, pandas-dev/pandas: Pandas, latest,  Zenodo,
  \dodoi{10.5281/zenodo.3509134}

\bibitem[{{Price-Whelan} {et~al.}(2018){Price-Whelan}, {Sip{\H{o}}cz},
  {G{\"u}nther}, {Lim}, {Crawford}, {Conseil}, {Shupe}, {Craig}, {Dencheva},
  {Ginsburg}, {VanderPlas}, {Bradley}, {P{\'e}rez-Su{\'a}rez}, {de Val-Borro},
  {Paper Contributors}, {Aldcroft}, {Cruz}, {Robitaille}, {Tollerud},
  {Coordination Committee}, {Ardelean}, {Babej}, {Bach}, {Bachetti}, {Bakanov},
  {Bamford}, {Barentsen}, {Barmby}, {Baumbach}, {Berry}, {Biscani}, {Boquien},
  {Bostroem}, {Bouma}, {Brammer}, {Bray}, {Breytenbach}, {Buddelmeijer},
  {Burke}, {Calderone}, {Cano Rodr{\'\i}guez}, {Cara}, {Cardoso}, {Cheedella},
  {Copin}, {Corrales}, {Crichton}, {D{\textquoteright}Avella}, {Deil},
  {Depagne}, {Dietrich}, {Donath}, {Droettboom}, {Earl}, {Erben}, {Fabbro},
  {Ferreira}, {Finethy}, {Fox}, {Garrison}, {Gibbons}, {Goldstein}, {Gommers},
  {Greco}, {Greenfield}, {Groener}, {Grollier}, {Hagen}, {Hirst}, {Homeier},
  {Horton}, {Hosseinzadeh}, {Hu}, {Hunkeler}, {Ivezi{\'c}}, {Jain}, {Jenness},
  {Kanarek}, {Kendrew}, {Kern}, {Kerzendorf}, {Khvalko}, {King}, {Kirkby},
  {Kulkarni}, {Kumar}, {Lee}, {Lenz}, {Littlefair}, {Ma}, {Macleod},
  {Mastropietro}, {McCully}, {Montagnac}, {Morris}, {Mueller}, {Mumford},
  {Muna}, {Murphy}, {Nelson}, {Nguyen}, {Ninan}, {N{\"o}the}, {Ogaz}, {Oh},
  {Parejko}, {Parley}, {Pascual}, {Patil}, {Patil}, {Plunkett}, {Prochaska},
  {Rastogi}, {Reddy Janga}, {Sabater}, {Sakurikar}, {Seifert}, {Sherbert},
  {Sherwood-Taylor}, {Shih}, {Sick}, {Silbiger}, {Singanamalla}, {Singer},
  {Sladen}, {Sooley}, {Sornarajah}, {Streicher}, {Teuben}, {Thomas},
  {Tremblay}, {Turner}, {Terr{\'o}n}, {van Kerkwijk}, {de la Vega}, {Watkins},
  {Weaver}, {Whitmore}, {Woillez}, {Zabalza}, \& {Contributors}}]{astropy:2018}
{Price-Whelan}, A.~M., {Sip{\H{o}}cz}, B.~M., {G{\"u}nther}, H.~M., {et~al.}
  2018, \aj, 156, 123, \dodoi{10.3847/1538-3881/aabc4f}

\bibitem[{{Rebull} {et~al.}(2017){Rebull}, {Stauffer}, {Hillenbrand}, {Cody},
  {Bouvier}, {Soderblom}, {Pinsonneault}, \& {Hebb}}]{Rebull2017}
{Rebull}, L.~M., {Stauffer}, J.~R., {Hillenbrand}, L.~A., {et~al.} 2017, \apj,
  839, 92, \dodoi{10.3847/1538-4357/aa6aa4}

\bibitem[{{Rebull} {et~al.}(2016){Rebull}, {Stauffer}, {Bouvier}, {Cody},
  {Hillenbrand}, {Soderblom}, {Valenti}, {Barrado}, {Bouy}, {Ciardi},
  {Pinsonneault}, {Stassun}, {Micela}, {Aigrain}, {Vrba}, {Somers},
  {Christiansen}, {Gillen}, \& {Collier Cameron}}]{Rebull2016}
{Rebull}, L.~M., {Stauffer}, J.~R., {Bouvier}, J., {et~al.} 2016, \aj, 152,
  113, \dodoi{10.3847/0004-6256/152/5/113}

\bibitem[{{Ricker} {et~al.}(2015){Ricker}, {Winn}, {Vanderspek}, {Latham},
  {Bakos}, {Bean}, {Berta-Thompson}, {Brown}, {Buchhave}, {Butler}, {Butler},
  {Chaplin}, {Charbonneau}, {Christensen-Dalsgaard}, {Clampin}, {Deming},
  {Doty}, {De Lee}, {Dressing}, {Dunham}, {Endl}, {Fressin}, {Ge}, {Henning},
  {Holman}, {Howard}, {Ida}, {Jenkins}, {Jernigan}, {Johnson}, {Kaltenegger},
  {Kawai}, {Kjeldsen}, {Laughlin}, {Levine}, {Lin}, {Lissauer}, {MacQueen},
  {Marcy}, {McCullough}, {Morton}, {Narita}, {Paegert}, {Palle}, {Pepe},
  {Pepper}, {Quirrenbach}, {Rinehart}, {Sasselov}, {Sato}, {Seager},
  {Sozzetti}, {Stassun}, {Sullivan}, {Szentgyorgyi}, {Torres}, {Udry}, \&
  {Villasenor}}]{TESS}
{Ricker}, G.~R., {Winn}, J.~N., {Vanderspek}, R., {et~al.} 2015, Journal of
  Astronomical Telescopes, Instruments, and Systems, 1, 014003,
  \dodoi{10.1117/1.JATIS.1.1.014003}

\bibitem[{{Rodr{\'\i}guez} {et~al.}(2016){Rodr{\'\i}guez},
  {Rodr{\'\i}guez-L{\'o}pez}, {L{\'o}pez-Gonz{\'a}lez}, {Amado}, {Ocando}, \&
  {Berdi{\~n}as}}]{Rodriguez2016}
{Rodr{\'\i}guez}, E., {Rodr{\'\i}guez-L{\'o}pez}, C., {L{\'o}pez-Gonz{\'a}lez},
  M.~J., {et~al.} 2016, \mnras, 457, 1851, \dodoi{10.1093/mnras/stw033}

\bibitem[{{Rousseeuw} \& {Croux}(1993)}]{Rousseeuw1993}
{Rousseeuw}, P.~J., \& {Croux}, C. 1993, Journal of the American Statistical
  Association, 88, 1273, \dodoi{10.1080/01621459.1993.10476408}

\bibitem[{{Santos} {et~al.}(2019){Santos}, {Garc{\'\i}a}, {Mathur}, {Bugnet},
  {van Saders}, {Metcalfe}, {Simonian}, \& {Pinsonneault}}]{Santos2019}
{Santos}, A.~R.~G., {Garc{\'\i}a}, R.~A., {Mathur}, S., {et~al.} 2019, \apjs,
  244, 21, \dodoi{10.3847/1538-4365/ab3b56}

\bibitem[{{Schatzman}(1962)}]{Schatzman1962}
{Schatzman}, E. 1962, Annales d'Astrophysique, 25, 18

\bibitem[{{Sellwood}(2014)}]{sellwood2014}
{Sellwood}, J.~A. 2014, Reviews of Modern Physics, 86, 1,
  \dodoi{10.1103/RevModPhys.86.1}

\bibitem[{{Sellwood} \& {Carlberg}(1984)}]{Sellwood1984}
{Sellwood}, J.~A., \& {Carlberg}, R.~G. 1984, \apj, 282, 61,
  \dodoi{10.1086/162176}

\bibitem[{{Silva Aguirre} {et~al.}(2017){Silva Aguirre}, {Lund}, {Antia},
  {Ball}, {Basu}, {Christensen-Dalsgaard}, {Lebreton}, {Reese}, {Verma},
  {Casagrande}, {Justesen}, {Mosumgaard}, {Chaplin}, {Bedding}, {Davies},
  {Handberg}, {Houdek}, {Huber}, {Kjeldsen}, {Latham}, {White}, {Coelho},
  {Miglio}, \& {Rendle}}]{Silva2017}
{Silva Aguirre}, V., {Lund}, M.~N., {Antia}, H.~M., {et~al.} 2017, \apj, 835,
  173, \dodoi{10.3847/1538-4357/835/2/173}

\bibitem[{{Simonian} {et~al.}(2019){Simonian}, {Pinsonneault}, \&
  {Terndrup}}]{simonian2019}
{Simonian}, G. V.~A., {Pinsonneault}, M.~H., \& {Terndrup}, D.~M. 2019, \apj,
  871, 174, \dodoi{10.3847/1538-4357/aaf97c}

\bibitem[{{Skumanich}(1972)}]{Skumanich1972}
{Skumanich}, A. 1972, \apj, 171, 565, \dodoi{10.1086/151310}

\bibitem[{{Soderblom}(2010)}]{Soderblom2010}
{Soderblom}, D.~R. 2010, \araa, 48, 581,
  \dodoi{10.1146/annurev-astro-081309-130806}

\bibitem[{{Spada} \& {Lanzafame}(2020)}]{Spada2020}
{Spada}, F., \& {Lanzafame}, A.~C. 2020, \aap, 636, A76,
  \dodoi{10.1051/0004-6361/201936384}

\bibitem[{{Spina} {et~al.}(2018){Spina}, {Mel{\'e}ndez}, {Karakas}, {dos
  Santos}, {Bedell}, {Asplund}, {Ram{\'\i}rez}, {Yong}, {Alves-Brito}, {Bean},
  \& {Dreizler}}]{Spina2018}
{Spina}, L., {Mel{\'e}ndez}, J., {Karakas}, A.~I., {et~al.} 2018, \mnras, 474,
  2580, \dodoi{10.1093/mnras/stx2938}

\bibitem[{{Spitzer} \& {Schwarzschild}(1951)}]{Spitzer1951}
{Spitzer}, Lyman, J., \& {Schwarzschild}, M. 1951, \apj, 114, 385,
  \dodoi{10.1086/145478}

\bibitem[{{Str{\"o}mberg}(1946)}]{Stromberg1946}
{Str{\"o}mberg}, G. 1946, \apj, 104, 12, \dodoi{10.1086/144830}

\bibitem[{{Ting} \& {Rix}(2019)}]{Ting2019}
{Ting}, Y.-S., \& {Rix}, H.-W. 2019, \apj, 878, 21,
  \dodoi{10.3847/1538-4357/ab1ea5}

\bibitem[{{van Saders} {et~al.}(2016){van Saders}, {Ceillier}, {Metcalfe},
  {Silva Aguirre}, {Pinsonneault}, {Garc{\'\i}a}, {Mathur}, \&
  {Davies}}]{vansaders2016}
{van Saders}, J.~L., {Ceillier}, T., {Metcalfe}, T.~S., {et~al.} 2016, \nat,
  529, 181, \dodoi{10.1038/nature16168}

\bibitem[{{van Saders} \& {Pinsonneault}(2013)}]{vansanders2013}
{van Saders}, J.~L., \& {Pinsonneault}, M.~H. 2013, \apj, 776, 67,
  \dodoi{10.1088/0004-637X/776/2/67}

\bibitem[{{van Saders} {et~al.}(2019){van Saders}, {Pinsonneault}, \&
  {Barbieri}}]{Saders2019}
{van Saders}, J.~L., {Pinsonneault}, M.~H., \& {Barbieri}, M. 2019, \apj, 872,
  128, \dodoi{10.3847/1538-4357/aafafe}

\bibitem[{{Veyette} \& {Muirhead}(2018)}]{Veyette2018}
{Veyette}, M.~J., \& {Muirhead}, P.~S. 2018, \apj, 863, 166,
  \dodoi{10.3847/1538-4357/aad40e}

\bibitem[{{Weber} \& {Davis}(1967)}]{Weber1967}
{Weber}, E.~J., \& {Davis}, Leverett, J. 1967, \apj, 148, 217,
  \dodoi{10.1086/149138}

\bibitem[{{Wenger} {et~al.}(2000){Wenger}, {Ochsenbein}, {Egret}, {Dubois},
  {Bonnarel}, {Borde}, {Genova}, {Jasniewicz}, {Lalo{\"e}}, {Lesteven}, \&
  {Monier}}]{simbad}
{Wenger}, M., {Ochsenbein}, F., {Egret}, D., {et~al.} 2000, \aaps, 143, 9,
  \dodoi{10.1051/aas:2000332}

\bibitem[{{W}es {M}c{K}inney(2010)}]{mckinney-proc-scipy-2010}
{W}es {M}c{K}inney. 2010, in {P}roceedings of the 9th {P}ython in {S}cience
  {C}onference, ed. {S}t\'efan van~der {W}alt \& {J}arrod {M}illman, 56 -- 61,
  \dodoi{10.25080/Majora-92bf1922-00a}

\bibitem[{{Wright} {et~al.}(2011){Wright}, {Drake}, {Mamajek}, \&
  {Henry}}]{Wright2011}
{Wright}, N.~J., {Drake}, J.~J., {Mamajek}, E.~E., \& {Henry}, G.~W. 2011,
  \apj, 743, 48, \dodoi{10.1088/0004-637X/743/1/48}

\bibitem[{{Xiang} {et~al.}(2019){Xiang}, {Ting}, {Rix}, {Sandford}, {Buder},
  {Lind}, {Liu}, {Shi}, \& {Zhang}}]{Xiang2019}
{Xiang}, M., {Ting}, Y.-S., {Rix}, H.-W., {et~al.} 2019, \apjs, 245, 34,
  \dodoi{10.3847/1538-4365/ab5364}

\bibitem[{{Yu} \& {Liu}(2018)}]{Yu2018}
{Yu}, J., \& {Liu}, C. 2018, \mnras, 475, 1093, \dodoi{10.1093/mnras/stx3204}

\end{thebibliography}
\bibliographystyle{aasjournal}



\end{document}